\documentstyle[12pt,hyperref]{article}
\newcommand{\be}[1]{ \begin{equation}\label{#1} }
\newcommand{\ee}{\end{equation}}
\newcommand{\bea}[1]{\begin{eqnarray}\label{#1} }
\newcommand{\eea}{\end{eqnarray}}

\newcommand{\II}{{\cal I}}

\newcommand{\FF}{{\cal F}}
\newcommand{\NN}{{\cal N}}
\newcommand{\p}{\partial}
\newcommand{\wt}{\widetilde}
\def\ZZZ{{\hskip-3pt\hbox{ Z\kern-1.6mm Z}}}
\def\zzz{{\hskip-3pt\hbox{ z\kern-1mm z}}}

\def\ZZZ{{\hbox{Z\kern-1.6mm Z}}}
\def\zzz{{\hbox{z\kern-1mm z}}}

\newcommand{\vp}{\varphi}

\newcommand{\vt}{\vartheta}

\newcommand{\ws}{{\wt\sigma}}
\newcommand{\wrh}{{\wt\rho}}
\newcommand{\wv}{{\wt v}}

\newcommand{\tI}{\wt\II}
\newcommand{\hI}{\wh\II}

\def\ZZZZ{{\hbox{Z\kern-2.1mm Z}}}

\newcommand{\MM}{{\cal M}}
\newcommand{\CC}{{\cal C}}

\newcommand{\OO}{{\cal O}}

\newcommand{\EE}{{\cal E}}
\newcommand{\LL}{{\cal L}}

\newcommand{\wh}{\widehat}
\newcommand{\wc}{\check}

\newcommand{\cp}{\wh\Phi}

\renewcommand{\wc}{\wt}

\newcommand{\ben}{\begin{eqnarray}\displaystyle}
\newcommand{\een}{\end{eqnarray}}
\newcommand{\refb}[1]{(\ref{#1})}
\newcommand{\sectiono}[1]{\section{#1}\setcounter{equation}{0}}

\def\one{{\hbox{ 1\kern-.8mm l}}}
\def\zero{{\hbox{ 0\kern-1.5mm 0}}}

\begin{document}

{}~
{}~

 \hfill\vbox{\hbox{hep-th/0609109}
}\break

\vskip .6cm

{\baselineskip20pt
\begin{center}
{\Large \bf
Dyon Spectrum in Generic $\NN=4$ Supersymmetric $\ZZZZ_N$
Orbifolds}

\end{center} }

\vskip .6cm
\medskip

\vspace*{4.0ex}

\centerline{\large \rm
Justin R. David, Dileep P. Jatkar and Ashoke Sen}

\vspace*{4.0ex}

\centerline{\large \it Harish-Chandra Research Institute}

\centerline{\large \it  Chhatnag Road, Jhusi,
Allahabad 211019, INDIA}

\vspace*{1.0ex}

\centerline{\it E-mail: 
justin,dileep,sen@mri.ernet.in,
ashoke.sen@cern.ch}

\vspace*{5.0ex}

\vskip 1in

\centerline{\bf Abstract} \bigskip

We find the exact spectrum of a class of quarter BPS dyons in a
generic $\NN=4$ supersymmetric $\ZZZ_N$ orbifold of type
IIA string theory on $K3\times T^2$ or $T^6$. 
We also find the asymptotic expansion
of the statistical entropy to first non-leading order in inverse power
of charges and show that it agrees with the entropy of a black hole
carrying same set of charges after taking into account the effect
of the four derivative Gauss-Bonnet term in the effective action
of the theory.

\vfill \eject

\baselineskip=18pt

\tableofcontents

\sectiono{Introduction and Summary} \label{sint}

We now have a good understanding of the spectrum of 1/4 BPS
states in a class of $\NN=4$ supersymmetric string theories which are
obtained as $\ZZZ_N$ orbifolds of type IIA string theory on
$K3\times T^2$ or $T^4\times T^2$ for prime values of 
$N$\cite{9607026,0412287,0505094,0506249,0508174,0510147,
0602254,0603066,0605210,0607155}. In each example studied
so far, the statistical entropy computed by taking the logarithm of the
degeneracy of states agrees with the entropy of the corresponding
black hole for large charges, not only in the leading order but also
in the first non-leading 
order\cite{0412287,0510147,0605210,0607155}. On the black hole
side this requires  taking into
account the effect of Gauss-Bonnet term in the low energy effective
action of the theory,
and use of Wald's generalized formula
for the black hole entropy in the presence of higher derivative
corrections\cite{9307038,9312023,9403028,9502009}.

In this paper we generalize this analysis to  $\NN=4$
supersymmetric theories, obtained as $\ZZZ_N$ orbifolds of
type IIA string theory on
$K3\times T^2$ or $T^4\times T^2$, for generic $N$ which are
not necessarily
prime. In this process we also demonstrate the relationship between
the black hole entropy and the statistical entropy in a more
explicit manner by comparing the expressions for various
coefficients rather than matching their final values.

Since the analysis of the paper involves a lot of technical details,
we shall summarize our results here. As in the case of 
\cite{0605210,0607155} we consider  
type IIB string theory on $\MM\times S^1\times \wt S^1$ where
$\MM$ is either K3 or $T^4$, and 
mod out this theory by a $\ZZZ_N$ symmetry
group generated by a transformation $g$ that involves
$1/N$ unit of shift along the circle $S^1$ 
together with an order
$N$ transformation
$\wt g$ in $\MM$. $\wt g$ is chosen in such a way that the final
theory has $\NN=4$ supersymmetry. We consider in this theory
 a configuration with a single D5-brane wrapped on 
 $\MM\times S^1$, $Q_1$ D1-branes wrapped on $S^1$, a single
 Kaluza-Klein monopole associated with the circle $\wt S^1$,
 $-n/N$ units of momentum along $S^1$ and $J$ units of momentum
 along $\wt S^1$\cite{0505094}.\footnote{Here all the units 
  refer to those in the orbifold
  theory. Thus for example if $S^1$ has radius $R$ then in the
  orbifold theory there will be periodicity under a translation by
  $2\pi R/N$ along $S^1$ together with an appropriate transformation
  on the rest of the conformal field theory. Hence the unit of momentum
  along $S^1$ is taken to be $N/R$.} 
 By making an S-duality transformation, followed
 by a T-duality along the circle $\wt S^1$ and a 
 six dimensional string-string duality, we can map this system to
 an asymmetric $\ZZZ_N$ orbifold of
 heterotic (for $\MM=K3$)\cite{9507050,9508144,9508154}
or type IIA (for $\MM=T^4$)\cite{9508064} string
 theory
 on $T^4\times S^1\times \wh S^1$, with $-n/N$ units of momentum
 along $S^1$, a single Kaluza-Klein monopole associated with 
 $\wh S^1$,
  $(Q_1-\beta)$ units of NS 5-brane charge along $T^4\times
  S^1$, $J$ units of NS 5-brane charge along 
  $T^4\times \wh S^1$
  and a single fundamental string wound along 
  $S^1$\cite{0605210}.
  Here $\beta$ is the Euler character of $\MM$ 
  divided by 24.
  If $Q_e$ and $Q_m$ denote the electric and magnetic charge
  vectors in this asymmetric orbifold description,\footnote{Unless
  mentioned otherwise, whenever we refer to electric or magnetic
  charges or T- or S-duality
  symmetry of the theory, we shall imply electric or magnetic
  charges or T- or S-duality symmetry in the
  asymmetric orbifold description.}
  and if $\cdot$
  denotes the T-duality invariant inner product in this description,
  then we have
 \be{eqdef}
 Q_e^2  \equiv Q_e\cdot Q_e
 = 2 n/N, \qquad Q_m^2  \equiv Q_m\cdot Q_m
 = 2 (Q_1- \beta   ), \qquad Q_e\cdot Q_m = J\, .
\ee
We denote by 
$d(Q_e,Q_m)$ the number of bosonic minus fermionic
quarter BPS supermultiplets carrying a given set of charges
$(Q_e, Q_m)$, a supermultiplet being considered bosonic
(fermionic) if it is obtained by tensoring the basic 64 dimensional
quarter BPS supermultiplet with a supersymmetry singlet  
bosonic (fermionic) state.
Our result for $d(Q_e,Q_m)$  is
\be{egg1int}
d(Q_e,Q_m) = {1\over N}\, \int _\CC d\wt\rho \, 
d\wt\sigma \,
d\wt v \, e^{-\pi i ( N\wt \rho Q_e^2
+ \wt \sigma Q_m^2/N +2\wt v Q_e\cdot Q_m)}\, {1
\over \wt\Phi(\wt \rho,\wt \sigma, \wt v)}\, ,
\ee
where $\wt\Phi$ is a function to be defined below and
$\CC$ is a three real dimensional subspace of the
three complex dimensional space labelled by $(\wt\rho,\ws,\wv)$,
given by
\bea{ep2kk}
Im\, \wt \rho=M_1, \quad Im \, \wt\sigma = M_2, \quad
Im \, \wt v = M_3, \nonumber \\
 0\le Re\, \wt\rho\le 1, \quad
0\le Re\, \wt\sigma\le N, \quad 0\le Re\,  \wt v\le 1\, ,
\een
where $M_1$, $M_2$ and $M_3$ are large but fixed positive
numbers.
Alternatively, we can express $d(Q_e,Q_m)$ as
\be{efo1}
d(Q_e, Q_m) = g\left({N\over 2} Q_e^2 , {1\over 2\, N}\, Q_m^2,
Q_e\cdot Q_m\right)\, ,
\ee
where $g(m,n,p)$ are
the coefficients of Fourier expansion of the function
$1/ \wt\Phi(\wt \rho,\wt \sigma, \wt v)$:
\be{efo2}
{1
\over \wt\Phi(\wt \rho,\wt \sigma, \wt v)}
=\sum_{m,n,p} g(m,n,p) \, e^{2\pi i (m\, \wt \rho + n\,
\wt\sigma
+ p\, \wt v)}\, .
\ee

In order to define $\wt\Phi$ we shall have to consider a 
2-dimensional
(4,4)
superconformal $\sigma$-model with target space $\MM$, modded
out by the 
$\wt\ZZZ_N$ group generated by the transformation $\wt g$
described earlier. In this theory we define\cite{9306096}
 \be{esi4akk}
F^{(r,s)}(\tau,z) \equiv {1\over N} Tr_{RR;\wt g^r} \left(\wt g^s
(-1)^{F_L+F_R}
e^{2\pi i \tau L_0} 
e^{-2\pi i \bar\tau \bar L_0}
e^{2\pi i F_L z}\right), \qquad 0\le r,s\le N-1\, ,
 \ee
where
$Tr$ denotes trace over all the Ramond-Ramond (RR) sector 
states twisted by $\wt g^r$ in the superconformal field
theory (SCFT) described above
before we project on to $\wt g$ invariant
states, $F_L$ and $F_R$ denote the world-sheet fermion 
numbers
associated with left and right chiral fermions in this SCFT and
$L_n$, $\bar L_n$ are the Virasoro generators in this SCFT with
additive factors of $-c_L/24$ and $-c_R/24$ included in the definitions
of $L_0$ and $\bar L_0$. In this convention the RR sector ground
state has $L_0=\bar L_0=0$.
$F^{(r,s)}(\tau,z)$ can be shown to have an expansion of the
form
\be{enewkk}
F^{(r,s)}(\tau,z) =\sum_{b=0}^1\sum_{j\in2\zzz+b, n\in \zzz/N
\atop
4n - j^2\ge -b^2} 
c^{(r,s)}_b(4n -j^2)
e^{2\pi i n\tau + 2\pi i jz}\, .
\ee
This defines the coefficients $c^{(r,s)}_b(u)$. We also define
\be{eqrsrevint}
Q_{r,s} = N\, 
\left( c^{(r,s)}_0(0)+ 2 \, c^{(r,s)}_1(-1)\right)\, ,
\ee
and
\be{edefap}
\wt \alpha={1\over 24N} \, Q_{0,0} - {1\over 2N}
\, \sum_{s=1}^{N-1} Q_{0,s}\, {e^{-2\pi i s/N}\over
(1-e^{-2\pi i s/N})^2 } \, 
, \qquad 
\wt \gamma= {1\over 24N} \, Q_{0,0}\, .
\ee
In terms of these coefficients the function $\wt\Phi$ appearing
in \refb{egg1int} is given by
\bea{edefwtphi}
&& \wt \Phi(\wt \rho,\wt \sigma,\wt v ) =
e^{2\pi i (\wt \alpha\wt\rho + \wt \gamma\ws 
+ \wt v)} \nonumber \\
&& \qquad \times \prod_{b=0}^1\, 
 \prod_{r=0}^{N-1}
\prod_{k'\in \zzz+{r\over N},l\in\zzz,j\in 2\zzz+b
\atop k',l\ge 0, j<0 \, {\rm for}
\, k'=l=0}
\left( 1 - \exp\left(2\pi i ( k'\wt \sigma   +  l\wt \rho +  j\wt v)
\right)\right)^{
\sum_{s=0}^{N-1} e^{-2\pi i sl/N } c^{(r,s)}_b(4k'l - j^2)} \, .
\nonumber \\
\eea
This expression for $\wt\Phi$, including the values
of $\wt\alpha$ and $\wt\gamma$,  reduces to the ones
studied earlier for prime values of 
$N$\cite{0510147,
0602254,0603066,0605210,0607155} except for an overall 
normalization factor. We have used a new normalization convention
for $\wt\Phi$ to simplify some of the formul\ae.

One point about the degeneracy formula given above is worth
mentioning. Eqs.\refb{egg1int} and \refb{efo1} are equivalent
only if the sum over $m$, $n$, $p$ in \refb{efo2}
are convergent for large
imaginary $\wrh$, $\ws$ and $\wv$. This in particular 
requires that for fixed $m$ and $n$
the sum over $p$ is bounded from below.  By examining the
formula \refb{edefwtphi} for $\wt\Phi$ and the fact that the
coefficients $c^{(r,s)}_b(u)$ are non-zero only for $4u\ge -b^2$,
we can verify that with the exception of the contribution from the
$k'=l=0$ term in this product, the other terms, when expanded
in a power series expansion in $e^{2\pi i\wrh}$ and $e^{2\pi i\ws}$,
does have the form of \refb{efo2} with $p$ bounded from below
for fixed $m$, $n$.
However for the $k'=l=0$ term, which arises from the
dynamics of the D1-D5 centre of mass motion in the Kaluza-Klein
monopole background and gives a contribution
$e^{-2\pi i \wv} / (1 - e^{-2\pi i \wv})^2$\cite{0605210}, 
there is an ambiguity in
carrying out the series expansion. We could either use the form given
above and expand the denominator in a series 
expansion in $e^{-2\pi i \wv}$, or express it in the form
$e^{2\pi i \wv} / (1 - e^{2\pi i \wv})^2$ and expand it in 
a series 
expansion in $e^{2\pi i \wv}$. It was shown in \cite{0605210} that
depending on the angle between $S^1$ and $\wt S^1$,
only one of these expansions produce the degeneracy formula
correctly. The physical spectrum actually changes as this angle
passes through $90^\circ$ since at this point the system is only
marginally stable.
On the other hand our degeneracy formula
\refb{egg1int} implicitly requires that we expand this factor in powers
of $e^{2\pi i \wv}$ since only in this case the sum over $p$ in
\refb{efo2} is bounded from below for fixed $m$, $n$.
Thus as it stands the formula is valid for a
specific range of values of the angle between $S^1$ and $\wt S^1$,
which, in the dual asymmetric orbifold
description of the system, corresponds to the sign of the axion field.
For the other sign of the axion we need to take $M_3$ to be large
and negative to get a correct formula for the degeneracy.

Another point about \refb{egg1int} is that it 
has been derived for special
charge vectors $Q_e$, $Q_m$ in a 
specific region of the moduli space,
-- the weakly coupling region in the original description as type
IIB string theory on $\MM\times S^1\times \wt S^1/\ZZZ_N$.
Thus although we have expressed the formula for $d(Q_e,Q_m)$
in a form that is independent of the asymptotic values of the
various moduli fields and as a function of the T-duality invariant
combinations $Q_e^2$, $Q_m^2$ and $Q_e\cdot Q_m$, it need
not have this form in all regions of the moduli space for all charge
vectors. In particular the spectrum could change discontinuously
across curves of marginal stability as we vary the 
moduli\cite{9712211}. Since the duality invariance of the theory 
only guarantees  that the spectrum remains invariant under a
simultaneous duality transformation of the moduli and the charge
vectors, we cannot invoke duality invariance to find $d(Q_e,Q_m)$
for general charge vectors unless we know the moduli dependence
of the formula from other sources.

S-duality invariance of the theory in the asymmetric orbifold
description corresponds to global diffeomorphism symmetry 
associated with
the torus $S^1\times \wt S^1$ in the original 
description of the theory. This leaves invariant the weak
coupling region of the theory, -- the region in which the
the degeneracy formula \refb{egg1int} has been derived. Thus
in this region the S-duality transformation should be a symmetry
of $d(Q_e, Q_m)$. The problem of verifying this directly however
is that we have derived eq.\refb{egg1int} for a specific choice of
the charge vectors $Q_e$, $Q_m$. If we assume that \refb{egg1int}
is valid for all charge vectors, -- at least in the weak coupling region,
--
then one can verify that this formula is indeed invariant under
S-duality transformation. Instead of taking this as a test of
S-duality transformation, -- which is expected to be true anyway, --
we can regard this as an indication that our formula
\refb{egg1int} for $d(Q_e, Q_m)$ is valid for general charge
vectors in the weak coupling region of the original theory.

By performing the integral over $\wt v$ in \refb{egg1int}
by picking up residues at
the poles of the integrand, and subsequent integral over $\wt\rho$, 
$\wt
\sigma$ by a saddle point approximation, we can extract the
behaviour of $d(Q_e,Q_m)$ for large charges. The result is that up
to first non-leading order, the entropy is given by extremizing a
statistical entropy function:
\be{enn18int}
-\wt\Gamma_B(\vec\tau) = {\pi\over 2 \tau_{2}} \, |Q_e +\tau Q_m|^2
- \ln g(\tau) -\ln g(-\bar\tau)
- (k+2) \ln (2\tau_{2}) + \hbox{constant} + \OO(Q^{-2})
\ee
with respect to real and imaginary parts of the complex variable
$\tau$. Here 
\be{ekint}
k={1\over 2}\, \sum_{s=0}^{N-1} \, c_0^{(0,s)}(0)
\ee
and
\be{enn13int}
g(\rho) = e^{2\pi i \wh\alpha\rho}\, 
\prod_{n=1}^\infty \prod_{r=0}^{N-1}
 \left( 1 - e^{2\pi i r/N}
e^{2\pi i n\rho}\right)^{s_{r}}\, ,
\ee
where
\be{enn14int}
s_{r}
= {1\over N} \sum_{s'=0}^{N-1}
e^{-2\pi i r s'/N} \, Q_{0,s'}\, , \qquad \wh\alpha = {1\over 24}
\, Q_{0,0}\, .
\ee
$|Q_e + \tau\, Q_m|^2$ appearing in \refb{enn18int} is to be
interpreted as
\be{extraeq}
Q_e^2 + 2\tau_1 \, Q_e\cdot Q_m + |\tau|^2 Q_m^2\, .
\ee

We can calculate the black hole entropy in this theory using the entropy
function formalism\cite{0506177,0508042}. The low energy effective
action is that of $\NN=4$ supergravity coupled to a certain number
of matter multiplets. However stringy corrections give rise to higher 
derivative terms in the action which include a Gauss-Bonnet term of
the form
\be{eh10-int}
\Delta\LL =  \phi(\tau,\bar \tau)\,
\left\{ R_{G\mu\nu\rho\sigma} R_G^{\mu\nu\rho\sigma}
- 4 R_{G\mu\nu} R_G^{\mu\nu}
+ R_G^2
\right\} \, ,
\ee
where $\tau$ denotes the complex structure modulus of the torus
$S^1\times \wt S^1$. The function $\phi(\tau,\bar\tau)$ can be
calculated using the method of \cite{9708062} and is given by
\be{eh10bint}
\phi(\tau,\bar \tau) = - {1\over 64\pi^2} \, \left( (k+2) \ln \tau_2 
+ \ln g(\tau) + \ln g(\bar \tau)\right)
+\hbox{constant}\, .
\ee
The entropy of a dyonic black hole, after taking into account
corrections due to the Gauss-Bonnet term, 
is given by the extremum of the black hole entropy
function\cite{0508042}
\be{eblackint}
\EE ={\pi\over 2 \tau_{2}} \, |Q_e +\tau Q_m|^2
- \ln g(\tau) -\ln g(-\bar\tau)
- (k+2) \ln (2\tau_{2}) + \hbox{constant} + \OO(Q^{-2})\, .
\ee
Comparing \refb{enn18int} and \refb{eblackint} we see that the black
hole entropy and the statistical entropy agree to this
order.\footnote{The full action contains
other four derivative terms besides the Gauss-Bonnet term and hence
there is no {\it a priori} justification for keeping only the Gauss-Bonnet
term in the effective action. However at least
for $Q_e^2>>Q_m^2, Q_e\cdot
Q_m$ when the coupling constant at the horizon in the asymmetric
orbifold description is small,
one can show that the Gauss-Bonnet term captures the effect
of complete set of four derivative
terms\cite{0506176,0508218,0608182,0609074}.
This is also true if we add to the action the set of all terms related
to the curvature squared term via supersymmetry 
transformation\cite{9906094,0007195}. Thus there is 
some non-renormalization theorem at work at least for $Q_e^2
>> Q_m^2, Q_e\cdot Q_m$. Our hope is that similar non-renormalization
theorems would also hold when all the charges are of the same order.}

The rest of the paper is organised as follows. 
Sections \ref{s0} and \ref{s3} contain general mathematical results
which will be useful for studies in the later sections.
In section \ref{s0} we study in detail some properties of the two
dimensional (4,4) superconformal field theory with target space
$\MM$ modded out by the group $\wt\ZZZ_N$ generated
by $\wt g$, and various
properties of the functions $F^{(r,s)}(\tau,z)$ and their
Fourier coefficients $c^{(r,s)}_b(u)$. Section  \ref{s3}
is devoted to studying various properties of the function $\wt\Phi(\wrh,
\ws,\wv)$ and some other related functions which are necessary for
studying the duality transformation properties as well as 
the asymptotic expansion of the statistical entropy.
Section \ref{sm}, which is the main section of this paper,
describes the computation of the degeneracy of dyons carrying a given
set of charges. 
As in the analysis of \cite{0605210,0607155} the contribution to the
dyon partition function comes from three separate
sources, -- the dynamics of the Kaluza-Klein monopole,
the overall motion of the D1-D5 system in the Kaluza-Klein monopole
background and the  motion of the D1-brane inside the D5-brane.
In section \ref{s3.5}
we prove the `S-duality' invariance of the degeneracy formula.  
Section \ref{s4} describes the asymptotic expansion 
of
the statistical entropy of the
system, defined as $\ln d(Q_e,Q_m)$, in the limit of large charges
up to first non-leading order. 
In section \ref{sblack} we calculate the entropy of a black
hole carrying the same charges by taking 
into account the Gauss-Bonnet term in the low energy effective action and
show that the result agrees with the statistical entropy to this order.

\sectiono{A Class of (4,4) Superconformal Field Theories} \label{s0}

In this section we shall introduce a class of (4,4) superconformal
field theories which will be useful for later analysis.

Let $\MM$ be either a $K3$ or a $T^4$ manifold, and let
$\wt g$ be an order $N$ discrete symmetry transformation 
acting on $\MM$.
We shall choose $\wt g$ in such a way that it 
satisfies the following properties (not all of which are independent):
\begin{enumerate}

\item We require that in an appropriate complex coordinate system
of $\MM$, $\wt g$  preserves the (0,2) and (2,0)
harmonic forms of $\MM$.

\item Let $\wt\ZZZ_N$ denote the group generated by $\wt g$.
We shall require that the orbifold   
$\wh\MM=\MM/\wt\ZZZ_N$ has
SU(2) holonomy.  

\item Let $\omega_i$ denote the harmonic
2-forms of $\MM$ and
\be{einter}
I_{ij}=\int_\MM  \omega_i \wedge \omega_j 
\ee
denote 
the intersection matrix of these
2-forms in $\MM$. When we 
diagonalize $I$ we get 3 eigenvalues $-1$ and a certain number
(say $P$) of the 
eigenvalues $+1$ ($P=19$ for $K3$ and 3 for $T^4$). We call
the 2-forms carrying eigenvalue $-1$  right-handed
2-forms and the 2-forms carrying eigenvalues $+1$ left-handed
2-forms. We shall choose $\wt g$ such that it
leaves invariant all the right-handed
2-forms. 

\item The $(4,4)$ superconformal field theory
with target space
$\MM$ has $SU(2)_L\times SU(2)_R$ R-symmetry group. 
We shall require that the transformation $\wt g$ 
commutes
with the (4,4) superconformal symmetry and the
$SU(2)_L\times SU(2)_R$ R-symmetry group of
the theory.
(For
$\MM=T^4$ the supersymmetry and the
R-symmetry groups are bigger, but $\wt g$ must be such that
only the (4,4) superconformal
symmetry and the 
$SU(2)_L\times SU(2)_R$ part of the R-symmetry group
commute with $\wt g$.)

\end{enumerate}

Let us now take an orbifold of this
(4,4) superconformal field theory by the
group $\wt \ZZZ_N$ generated by the transformation $\wt g$, and
define\cite{9306096}
 \be{esi4aint}
F^{(r,s)}(\tau,z) \equiv {1\over N} Tr_{RR;\wt g^r} \left(\wt g^s
(-1)^{F_L+F_R}
e^{2\pi i \tau L_0} 
e^{-2\pi i \bar\tau \bar L_0}
e^{2\pi i F_L z}\right), \qquad 0\le r,s\le N-1\, ,
 \ee
where
$Tr$ denotes trace over all the Ramond-Ramond (RR) sector 
states twisted by $\wt g^r$ in the SCFT described above
before we project on to $\wt g$ invariant
states, $L_n$, $\bar L_n$ denote the left- and right-moving
Virasoro generators and $F_L$ and $F_R$ 
denote the world-sheet fermion 
numbers
associated with left and right-moving sectors in this SCFT.
Equivalently we can identify $F_L$ ($F_R$) as twice
the  generator 
of the $U(1)_L$ ($U(1)_R$) subgroup of the
$SU(2)_L\times SU(2)_R$
R-symmetry group of this conformal field 
theory.\footnote{At this stage we are describing an abstract
conformal field theory without connecting it to string theory.
In all cases where we use this conformal field theory
to describe a fundamental string world-sheet theory
or world-volume theory
of some soliton, we 
shall use the Green-Schwarz formulation. Thus the world-sheet
fermion number of this SCFT
will represent the space-time fermion number in string theory.}
As in 
\cite{0602254} we include in the definition
of $L_0$, $\bar L_0$ additive factors of $-c_L/24$ and $-c_R/24$
respectively, so that RR sector ground state has $L_0=\bar L_0=0$.
Due to the insertion of $(-1)^{F_R}$ factor in the trace the
contribution to $F^{(r,s)}$ comes only from the $\bar L_0=0$
states. As a result $F^{(r,s)}$ does not depend on $\bar\tau$.

For $\wt g$ satisfying the conditions described earlier
the functions $F^{(r,s)}(\tau,z)$ have the form
\be{efhrel}
F^{(r,s)}(\tau,z) = h_0^{(r,s)}(\tau) \, \vt_3(2\tau,2z) +
h_1^{(r,s)}(\tau) \, \vt_2(2\tau,2z)\, .
\ee
This follows from the fact that $\vt_3(2\tau,2z)$ and $\vt_2(2\tau,2z)$
are the  characters of the $SU(2)_L$ level 1 current algebra which
is a symmetry of this SCFT. The functions 
$h_b^{(r,s)}(\tau)$ in turn have
expansions of the form
\be{ehbexp}
h_b^{(r,s)}(\tau) = \sum_{n\in{1\over N}\zzz -{b^2\over 4}}
c_b^{(r,s)}(4n) e^{2\pi i n\tau}\, .
\ee
This defines the coefficients $c^{(r,s)}_b(u)$. We shall justify
the restriction on the allowed values of $n$ shortly.
Using the known expansion of $\vt_3$ and $\vt_2$:
\be{eg1}
\vartheta_3(2\tau, 2 z) = \sum_{j\in 2\zzz} e^{2\pi i j z} e^{\pi i
\tau j^2/2}, 
\qquad \vartheta_2(2\tau, 2 z) = \sum_{j\in 2\zzz+{1}} e^{2\pi i j z} 
e^{\pi i \tau j^2/2},
\ee
 we get
\be{enewint}
F^{(r,s)}(\tau,z) =\sum_{b=0}^1\sum_{j\in2\zzz+b, n\in \zzz/N} 
c^{(r,s)}_b(4n -j^2)
e^{2\pi i n\tau + 2\pi i jz}\, .
\ee
Since in the RR sector the  $L_0$ eigenvalue is $\ge 0$ for any
state, 
it follows from \refb{efhrel}-\refb{eg1} that
\be{evanish}
c^{(r,s)}_0(u)=0 \quad \hbox{for $u<0$},
\qquad c^{(r,s)}_1(u)=0 \quad \hbox{for $u<-1$}\, .
\ee

$F^{(r,s)}(\tau,z)$ defined in \refb{esi4aint}
may be regarded as the partition function 
on a torus with modular parameter
$\tau$ with $\wt g^s \, e^{2\pi i F_L z}$ twist along the
$b$-cycle and $\wt g^r$ twist along the $a$-cycle.
If $(\sigma_1,\sigma_2)$ denote the coordinates of this torus, each
with period 1, then
under $\sigma_1\to -\sigma_1$, $\sigma_2\to -\sigma_2$, the
quantum numbers $r$ and $s$ change sign and also
$z\to -z$. Thus $F^{(r,s)}(\tau,z)=F^{(-r,-s)}(\tau, -z)$.
It then follows
from \refb{ehbexp}, \refb{enewint},  that
\be{ecrev}
h^{(r,s)}_b(\tau)
= h^{(-r,-s)}_b(\tau)\, , \qquad c^{(r,s)}_b(u) = c^{(-r,-s)}_b(u)\, .
\ee
Furthermore,
since under $(\sigma_1,\sigma_2)\to (\sigma_1+\sigma_2,\sigma_2)$
the modular parameter $\tau$ gets shifted by 1 and $(r,s)\to (r, s+r)$,
we must have $F^{(r,s+r)}(\tau+1,z)=F^{(r,s)}(\tau, z)$. Since
$(r,s)$ are defined modulo $N$ we get $F^{(r,s)}(\tau, z)
=F^{(r,s)}(\tau+N, z)$. This is the physical origin of the 
restriction $n\in \ZZZ/N$ in \refb{enewint} and $n\in \ZZZ/N-b^2/4$
in \refb{ehbexp}.

The $n=0$ terms in the expansion \refb{enewint}
is given by the contribution to \refb{esi4aint} 
from the RR sector
states with $L_0=\bar L_0=0$.
For $r=0$, \i.e.\ in the untwisted
sector, these states
are in one to one correspondence with 
harmonic $(p,q)$ forms on $\MM$, with
$(p-1)$ and $(q-1)$ measuring the quantum numbers
$F_L$ and $F_R$\cite{wittop,vafatop}. 
Thus $N\, c^{(0,s)}_0(0)$, being $N\times$
the coefficient of the $n=0$, $j=0$ term in \refb{enewint},
measures the number
of harmonic $(1,q)$ forms weighted by $(-1)^{q-1} \wt g^s$,
and $N\, c^{(0,s)}_1(-1)$, being $N\times$
the coefficient of the $n=0$,
$j=-1$ (or $j=1$) term in \refb{enewint}, 
measures the number of harmonic
$(0,q)$ (or $(2,q))$ forms weighted by $(-1)^{q}\wt g^s$.
If $\MM=K3$ then the only $(0,q)$ forms are $(0,0)$ and $(0,2)$
forms both of which are invariant under $\wt g$. Thus we have
\be{eck3}
c^{(0,s)}_1(-1) ={2\over N} \quad \hbox{for $\MM=K3$}\, .
\ee
On the other hand for $\MM=T^4$
one can represent the explicit action of $\wt g$
in an appropriate complex coordinate system $(z^1,z^2)$ as
\be{eoneform}
dz^1 \to e^{2\pi i/N} dz^1, \quad dz^2\to e^{-2\pi i/N} dz^2\, ,
\quad d\bar z^1 \to e^{-2\pi i/N} d\bar
z^1, \quad d\bar z^2\to e^{2\pi i/N} d\bar z^2\, ,
\ee
so that it preserves the $(2,0)$ and $(0,2)$ forms $dz^1\wedge dz^2$
and $d\bar z^1\wedge d\bar z^2$.
Using this one can  work out 
its action on all the 2-, 3- and 4-forms:
\bea{etwoform}
&&dz^1\wedge dz^2\to dz^1\wedge dz^2, \quad
dz^1\wedge d\bar z^1\to dz^1\wedge d\bar z^1, \quad
dz^1\wedge d\bar z^2\to e^{4\pi i/N} \, dz^1\wedge d\bar z^2,
\nonumber \\
&&d\bar z^1\wedge d\bar z^2\to d\bar z^1\wedge d\bar z^2, \quad
dz^2\wedge d\bar z^2\to dz^2\wedge d\bar z^2, \quad
d\bar z^1\wedge d z^2\to e^{-4\pi i/N} \, d\bar z^1\wedge d z^2\, ,
\nonumber \\
\eea
\bea{ethreeform}
&& dz^1\wedge dz^2\wedge d\bar z^1\to e^{-2\pi i/N}\,
dz^1\wedge dz^2\wedge d\bar z^1, \quad
dz^1\wedge dz^2\wedge d\bar z^2\to e^{2\pi i/N}\,
dz^1\wedge dz^2\wedge d\bar z^2, \nonumber \\
&& d\bar z^1\wedge d\bar z^2\wedge d  z^1\to e^{2\pi i/N}\,
d\bar z^1\wedge d\bar z^2\wedge d  z^1 , \quad
d\bar z^1\wedge d\bar z^2\wedge d  z^2\to e^{-2\pi i/N}\,
d\bar z^1\wedge d\bar z^2\wedge d  z^2, \nonumber \\
\eea
\be{efourform}
dz^1\wedge dz^2\wedge d\bar z^1\wedge d\bar z^2 \to
dz^1\wedge dz^2\wedge d\bar z^1\wedge d\bar z^2\, .
\ee
This shows that
the $(0,0)$ and $(0,2)$ forms are
invariant under $\wt g$ but the two $(0,1)$ forms carry 
$\wt g$ eigenvalues $\pm 2\pi/N$. Thus we have
\be{ect4}
c^{(0,s)}_1(-1) ={1\over N} 
\left(2 - e^{2\pi is/N} - e^{-2\pi is/N}\right) \quad 
\hbox{for $\MM=T^4$}\, .
\ee
\refb{etwoform} also shows that  
$\wt g$ acts trivially on four of the 2-forms, and acts as a
rotation by $4\pi/N$ in the two dimensional subspace spanned by
the other two 2-forms. By writing the
2-forms in the real basis one can easily verify that the 2-forms
which transform non-trivially under $\wt g$ correspond to 
left-handed 2-forms. 
These results will be useful later.

Another useful set of results emerges by taking the $z\to 0$
limit of eqs.\refb{esi4aint} and \refb{enewint}. This gives
\be{efqrs}
\sum_{b=0}^1\sum_{j\in2\zzz+b, n\in \zzz/N} 
c^{(r,s)}_b(4n -j^2)
e^{2\pi i n\tau} = {1\over N}\, Q_{r,s}\, ,
\ee
where
\be{eqrs}
Q_{r,s}= Tr_{RR;\wt g^r} \left(\wt g^s
(-1)^{F_L+F_R}
e^{2\pi i \tau L_0} e^{-2\pi i \bar\tau \bar L_0}
\right), \qquad 0\le r,s\le N-1\, .
 \ee
$Q_{r,s}$ is independent of $\tau$ and $\bar\tau$ since the 
$(-1)^{F_L+F_R}$
insertion in the trace makes the contribution from the $(L_0,\bar L_0)
\ne 
(0,0)$
states cancel. Thus \refb{efqrs} gives
\be{ecqrel}
\sum_{b=0}^1\sum_{j\in2\zzz+b} 
c^{(r,s)}_b(4n -j^2) = {1\over N} \, Q_{r,s}\, \delta_{n,0}\, .
\ee
Setting $n=0$ in the above equation and using eq.\refb{evanish} we get
\be{eqrsrev}
Q_{r,s} = N\, 
\left( c^{(r,s)}_0(0)+ 2 \, c^{(r,s)}_1(-1)\right)\, .
\ee
For $r=0$, \i.e.\ in the untwisted sector, the trace in \refb{eqrs}
reduces to a sum over the harmonic  
forms of $\MM$. Since $F_L+F_R$ is mapped to the degree of
the harmonic form, $Q_{0,s}$ has the interpretation of trace of
$(-1)^p\wt g^s$ over the harmonic $p$-forms of $\MM$. In 
particular we have
\be{eeuler}
Q_{0,0}= \chi(\MM)\, ,
\ee
where $\chi(\MM)$ denotes the Euler number of $\MM$.

For later use we shall define
\be{ewhfrs}
\wh F^{(r,s)}(\tau,z)
= {1\over N}\,
\sum_{s'=0}^{N-1}\sum_{r'=0}^{N-1}
e^{-2\pi i rs'/N} e^{2\pi i r's/N} \, F^{(r',s')}(\tau, z)\, .
\ee
$\wh F^{(r,s)}(\tau,z)$ satisfies properties similar to that of 
$F^{(r,s)}(\tau, z)$. In particular we have the relations:
\be{enewintwh}
\wh F^{(r,s)}(\tau,z) =\sum_{b=0}^1\sum_{j\in2\zzz+b, n\in \zzz/N} 
\wh c^{(r,s)}_b(4n -j^2)
e^{2\pi i n\tau + 2\pi i jz}\, ,
\ee
where 
\be{enn9e}
\wh c^{(r,s)}_b(u) = {1\over N}\, \sum_{r'=0}^{N-1}
\sum_{s'=0}^{N-1}\, e^{2\pi i (sr'-rs')/N} c^{(r',s')}_b(u)\, .
\ee
We also
have the analog of eq.\refb{ecqrel}
\be{efqrswh}
\sum_{b=0}^1\sum_{j\in2\zzz+b} 
\wh c^{(r,s)}_b(4n -j^2)
e^{2\pi i n\tau} = {1\over N}\, \wh Q_{r,s}\, \delta_{n,0}\, ,
\ee
where
\be{eqwhq}
\wh Q_{r,s} = {1\over N}\,
\sum_{s'=0}^{N-1}\sum_{r'=0}^{N-1} 
\, e^{2\pi i (sr'-rs')/N} \, Q_{r',s'}\, .
\ee

\sectiono{Siegel Modular Forms from Threshold Integrals} \label{s3}

In this section we shall prove various properties of $\wt\Phi$ defined
in \refb{edefwtphi} by
relating it to a `threshold integral'\cite{9512046}. 
We begin 
by defining:
\be{edefomega}
\Omega=\pmatrix{\rho  & v \cr v  & \sigma}\, ,
\ee
and
\bea{e7n}
{1\over 2} p_R^2 &=& {1\over 4 \det Im  \Omega} |-m_1 \rho  +
m_2 + n_1 \sigma + n_2 (\sigma\rho -v^2) + j v |^2, \nonumber \\
{1\over 2} p_L^2
&=& {1\over 2}  p_R^2 + m_1 n_1 + m_2 n_2 + {1\over 4} j^2\, ,
\eea
where $\rho$, $\sigma$ and $v$ are three
complex variables. We now consider the `threshold integrals'
\be{rwthrint}
\tI(\rho , \sigma, v ) = \sum_ {r, s =0}^{N-1}\sum_{b=0}^1
\tI_{r, s, b}\, , \qquad 
\hI(\rho , \sigma, v)
= \sum_ {r, s =0}^{N-1}\sum_{b=0}^1\, \hI_{r,s,b}\, ,
\ee
where
\be{defirsl}
\tI_{r,s,b} = \int_{\FF} \frac{d^2\tau}{\tau_2} 
\sum_{\stackrel{m_1, m_2, n_2 \in \zzz}{n_1\in \zzz+ \frac{r}{N}, 
j\in 2\zzz + b}}
q^{p_L^2/2} \bar q^{ p_R^2/2} e^{2\pi i  m_1 s/N}
h_b^{(r,s)}(\tau)\, ,
\ee
and
\be{enn6}
\hI_{r,s,b} = \int_{\FF} \frac{d^2\tau}{\tau_2} 
\sum_{\stackrel{m_1, n_1\in \zzz, m_2\in \zzz/N}{n_2\in N\zzz+ r, 
j\in 2\zzz + b}}
q^{p_L^2/2} \bar q^{ p_R^2/2} e^{2\pi i m_2 s} 
h_b^{(r,s)}(\tau)\, ,
\ee
with
\be{edefq}
q\equiv e^{2\pi i\tau}\, .
\ee
Let us now introduce another
set of variables 
$(\wc\rho,\wc\sigma,\wc v)$ related
to $(\rho,\sigma, v)$ via the relations
\be{e6na}
   \wc\rho={1\over N}\, 
   {1\over 2v-\rho-\sigma}, \qquad
   \wc\sigma = N\,
   {v^2-\rho\sigma \over 2v-\rho-\sigma}, \qquad
    \wc v =
   {v-\rho \over 2v-\rho-\sigma}\, ,
\ee
or equivalently,
\be{e5n}
\rho 
   = {\wc \rho \wc\sigma - \wc v^2\over N\wc\rho}, 
   \qquad \sigma = {\wc\rho \wc \sigma - (\wc v - 1)^2\over  
   N\wc\rho}, \qquad
   v 
=   {\wc\rho \wc\sigma - \wc v^2 + \wc v\over N\wc\rho}\, .
\ee
We also define
\be{edefomega2}
\wc\Omega=\pmatrix{\wc\rho  & \wc v \cr \wc v  & \wc \sigma}\, .
\ee
By relabelling the indices $m_1$, $m_2$, $n_1$, $n_2$ in
eqs.\refb{defirsl}-\refb{enn6} one can easily prove the relations
\be{enn4}
\hI(\rho,\sigma,v)=\tI(\wc\rho,
\wc\sigma,\wc v)\, .
\ee
In the same way one can show that under a transformation of the
form
\be{etr1}
 \Omega\to (A \Omega+B)(C \Omega+D)^{-1}\, ,
\ee
$\hI(\rho,\sigma, v)$ remains invariant for the following
choices of the matrices $A$, $B$, $C$, $D$:
\ben \label{egroup}
   \pmatrix{A&B\cr C&D}  &=& \pmatrix{ a & 0 & b & 0 \cr
     0 & 1 & 0 & 0\cr c & 0 & d & 0\cr 0 & 0 & 0 & 1}\, ,
   \qquad ad-bc=1, \quad \hbox{$c=0$ mod $N$, \quad $a,d=1$
     mod $N$}
   \nonumber \\
   \pmatrix{A&B\cr C&D}  &=& 
   \pmatrix{0 & 1 & 0 & 0 \cr -1 & 0 & 0 & 0\cr
     0 & 0 & 0 & 1\cr 0 & 0 & -1 & 0}\, , \nonumber \\
   \pmatrix{A&B\cr C&D}  &=& 
   \pmatrix{ 1 & 0 & 0 & \mu \cr
     \lambda & 1 & \mu & 0\cr 0 & 0 & 1 & -\lambda\cr
     0 & 0 & 0 & 1}\, , \qquad \lambda, \mu \in \ZZZ.
   \nonumber \\
\een
The group of transformations generated by these matrices is a subgroup
of the Siegel modular group $Sp(2,\ZZZ)$; 
we shall denote this subgroup by
$\wh G$.\footnote{For prime values of $N$ the group $\wh G$ 
is identical to the group $G$ introduced in \cite{0510147}.} Via 
eq.\refb{enn4} this also induces a
group of symmetry transformations of $\tI(\wrh,\ws,\wv)$;
we shall denote this group by
$\wt G$.

We can now follow the procedure of \cite{0602254} to evaluate the
integrals $\tI$ and $\hI$. Since the procedure is identical to
that in \cite{0602254}, we shall only quote the final results:
\be{enn7}
\tI(\rho ,\sigma,v ) = -2 \ln \left[
 ( \det\,{\rm Im}\Omega)^k \right] - 2\ln \wt\Phi (\rho ,\sigma, v )
- 2\ln \bar{\wt\Phi}(\rho ,\sigma, v ) +\hbox{constant}
\ee
and
\be{enn9}
\hI(\rho ,\sigma, v ) = -2 \ln \left[
 ( \det\,{\rm Im}\Omega)^k \right] - 2\ln \cp (\rho ,\sigma,v )
- 2\ln \bar{\cp}(\rho ,\sigma,v ) +\hbox{constant}
\ee
where
\be{ekvalue}
k={1\over 2}\, \sum_{s=0}^{N-1} \, c_0^{(0,s)}(0)\, ,
\ee
\bea{enn9a}
&& \wt \Phi(\rho ,\sigma, v ) =  
e^{2\pi i (\wt \alpha \rho + \wt \gamma\sigma 
+   v)} \nonumber \\
&& \qquad \times \prod_{b=0}^1\, 
 \prod_{r=0}^{N-1}
\prod_{k'\in \zzz+{r\over N},l\in\zzz,j\in 2\zzz+b
\atop k',l\ge 0, j<0 \, {\rm for}
\, k'=l=0}
\left( 1 - e^{2\pi i ( k'  \sigma   +  l  \rho +  j  v)
}\right)^{
\sum_{s=0}^{N-1} e^{-2\pi i sl/N } c^{(r,s)}_b(4k'l - j^2)} 
\nonumber \\
\eea
and
\bea{enn9c}
\cp(\rho,\sigma,v) &=&  e^{2\pi i \left(
\wh \alpha\rho+\wh \gamma\sigma+\wh\beta v\right) } \nonumber \\
&& \prod_{b=0}^1\,
\prod_{r,s=0}^{N-1}\,  \prod_{(k',l)\in \zzz,j\in 2\zzz+b\atop
k',l\ge 0, j<0 \, {\rm for}
\, k'=l=0}
\Big\{ 1 - e^{2\pi i r / N} \, e^{ 2\pi i ( k' \sigma + l \rho + j v) 
}
\Big\}^{ \wh c^{(r,s)}_b(4k'l - j^2)
}  \nonumber \\
\eea
with $\wh c^{(r,s)}_b(u)$ given in \refb{enn9e}, and
\bea{enn9d}
&& \wt \alpha={1\over 24N} \, Q_{0,0} - {1\over 2N}
\, \sum_{s=1}^{N-1} Q_{0,s}\, {e^{-2\pi i s/N}\over
(1-e^{-2\pi i s/N})^2 } \, 
, \qquad 
\wt \gamma= {1\over 24N} \, Q_{0,0}, \nonumber \\
&&    \wh \alpha= \wh\beta = \wh \gamma 
= {1\over 24} 
\, Q_{0,0}={1\over 24} 
\, \chi(\MM)\, .
\eea
The quantities $Q_{r,s}$ have been defined in 
eqs.\refb{eqrs}. 
In arriving at
\refb{enn9a}-\refb{enn9c} we have used the relations
\refb{ecrev}, \refb{ecqrel} and also \refb{eck3}, \refb{ect4}.
The constant $k$ defined in \refb{ekvalue} 
has the interpretation of being
half the number of $\wt g$ invariant $(1,q)$ forms weighted by
$(-1)^{q+1}$.

It now follows from  \refb{enn4},
\refb{enn7} and \refb{enn9} that
\be{enn11}
\wt\Phi(\wc\rho,\wc\sigma,\wc v)=C_1\, 
(2v -\rho-\sigma)^k\, \cp(\rho,\sigma,v)
\ee
where $C_1$ is a constant. Furthermore given the
invariance of  $\tI$ and $\hI$ under the groups  $\wt G$
and $\wh G$, it follows that  $\wt\Phi$ and $\cp$
transform as modular forms of weight $k$ under the groups
$\wt G$ and $\wh G$ respectively.

{}From \refb{enn9c}, \refb{eck3}, \refb{ect4} and \refb{efqrswh}
 it is easy to see that for small $v$
\be{enn12}
\wh\Phi(\rho, \sigma,  v)=   -4\pi^2 \, v^2 \, g( \rho)\, 
g( \sigma) + \OO(  v^4)
\ee
where
\be{enn13}
g(\rho) = e^{2\pi i \wh\alpha\rho}\,
\prod_{n=1}^\infty \prod_{r=0}^{N-1}
 \left( 1 - e^{2\pi i r/N}
e^{2\pi i n\rho}\right)^{s_{r}}\, ,
\ee
\be{enn14}
s_{r} = {1\over N}\, \sum_{s=0}^{N-1}\, \wh Q_{r,s} 
= {1\over N} \sum_{s'=0}^{N-1}
e^{-2\pi i r s'/N} \, Q_{0,s'}\,.
\ee
Eq.\refb{enn11} then gives, for small $v$, \i.e.\ small 
$\wc\rho\wc\sigma-\wc v^2+\wc v$,
\be{enn15}
\wt\Phi(\wc\rho,\wc\sigma,\wc v)=-4\pi^2\,  C_1\, 
(2v -\rho-\sigma)^k\, v^2 g(\rho) \, g(\sigma) + \OO(v^4)\, .
\ee
$s_r$ has the interpretation of being the number of harmonic
$p$-forms in $\MM$ with $\wt g$ eigenvalue $e^{2\pi i r/N}$
weighted by $(-1)^p$. Thus it is an integer.

We can determine the locations of the other zeroes and poles
of $\wt\Phi(\wrh,\ws,\wv)$ by identifying the 
logarithmic singularities
of $\tI(\wrh,\ws,\wv)$ as in 
\cite{0605210}. One finds that
$\wt\Phi(\wrh ,\ws,\wv )$ has possible zeroes at
\bea{ev4.a}
&&  \left( n_2 ( \ws\wrh -\wv ^2) + j\wv   
+ n_1 \ws  - \wrh  m_1 + m_2
\right)= 0\nonumber \\
&& \quad  \hbox{for $m_1,m_2,n_2\in \ZZZ$, 
$n_1\in{1\over N}\ZZZ$,
$j\in 2\ZZZ+1$, $m_1n_1+m_2n_2+{j^2\over 4}={1\over 4}$}\,  .
\een
The order of the zero is given by
\be{eor1}
\sum_{s=0}^{N-1}   e^{2\pi i m_1 s / N} c^{(r,s)}_1(-1), \qquad 
\hbox{$r=N\, n_1$ mod $N$}\, .
\ee
For $N\ge 5$ there are additional possible
zeroes of $\wt \Phi(\wt\rho,\ws,\wv)$
at
\bea{eaddz}
&& \left( n_2 ( \ws\wrh -\wv ^2) + j\wv   
+ n_1 \ws  - \wrh  m_1 + m_2
\right)= 0 \nonumber \\
&& \quad  \hbox{for $m_1,m_2,n_2\in \ZZZ$, 
$n_1\in{1\over N}\ZZZ$,
$j\in 2\ZZZ+1$, $m_1n_1+m_2n_2+{j^2\over 4}=
{1\over 4}-{1\over N}$}\,  .
\eea
The order of the zero is
\be{eex2}
\sum_{s=0}^{N-1}
e^{2\pi i m_1 s / N}\, c_1^{(r,s)}(-1+{4\over N})\, , \qquad 
\hbox{$r=Nn_1$ mod $N$}\,.
\ee
\refb{eor1} has the interpretation as the number of $\wt g^r$ twisted
states with $\wt g$ eigenvalue $e^{-2\pi i m_1/N}$, $F_L=1$ (or $F_L
=-1$) and $L_0=\bar L_0=0$, weighted by $(-1)^{F_L+F_R}$.
\refb{eex2} has the interpretation as the number of $\wt g^r$ twisted
states with $\wt g$ eigenvalue $e^{-2\pi i m_1/N}$, $F_L=1$ (or $F_L
=-1$), $L_0=1/N$ and $\bar L_0=0$, weighted by $(-1)^{F_L+F_R}$.
Thus both numbers are integers.

\sectiono{Dyon Partition Function} \label{sm}

We now
consider type IIB string theory compactified on $\MM\times
S^1\times \wt S^1$, $\MM$ being either K3 or $T^4$.
For definiteness we shall label $S^1$ and $\wt S^1$ by coordinates
with period $2\pi$. We then
take an orbifold of this theory
by a  discrete $\ZZZ_N$ transformation
generated by a transformation $g$, where $g$ involves 
a $2\pi/N$ translation along $S^1$ together with an order
$N$ transformation
$\wt g$ on $\MM$ described in section \ref{s0}.
Due to the properties of $\wt g$ described earlier, the 
resulting orbifold 
preserves all the supersymmetries of type IIB string
theory compactified on $K3\times S^1\times \wt S^1$. 
Thus if $\MM$ is $T^4$ then
the orbifolding  breaks half of the supersymmetries whereas for
$\MM=K3$ the orbifolding preserves all the supersymmetries.
By making an S-duality transformation of type IIB string theory, 
followed by a T-duality transformation
on the circle
$\wt S^1$ and a string-string duality transformation relating type
IIA string theory on $\MM$ to type IIA or heterotic string theory
on $T^4$, one can obtain
a dual description of these theories as  asymmetric orbifolds
of 
heterotic on $T^6$ for $\MM=K3$ and
asymmetric orbifolds of type IIB on $T^6$ for $\MM=T^4$.
In this description 
all
the space-time supersymmetries arise from the right-moving
sector of the fundamental string 
world-sheet\cite{0605210,0607155}. We shall choose the coordinates
along the circles $S^1$ and $\wt S^1$ such that before the orbifold
projection they have periodicity $2\pi$.

In the original description of the theory as type IIB on
$(\MM\times S^1\times \wt S^1)/\ZZZ_N$ we
consider
a system containing a single D5-brane
wrapped on $\MM\times S^1/\ZZZ_N$, $Q_1$ 
D1-branes wrapped on
$S^1/\ZZZ_N$, momentum $-n$ along $S^1$, 
momentum $J$ along $\wt S^1$ and a Kaluza-Klein monopole
associated with the compact circle $\wt S^1$.
In the dual asymmetric orbifold
description, the quantum numbers $n$ and the single
Kaluza-Klein monopole charge in the original theory appear as
momentum $-n$ and single fundamental string wound along $S^1$.
Hence they form
part of the electric charge vector $Q_e$. On the other hand
the D1-brane, D5-brane and the momentum along
$\wt S^1$ in the original theory correspond to a single Kaluza-Klein
monopole and $(Q_1-\beta)$ H-monopoles associated with the
dual circle of $\wt S^1$, and $J$ H-monopoles associated with
the circle $S^1/\ZZZ_N$ in the dual theory, where
\be{ed2}
\beta = {1\over 24} \chi(\MM)\, ,
\ee
$\chi(\MM)$ being the Euler character of $\MM$. Thus they form
part of the magnetic
charge vector $Q_m$, and we have\cite{0605210,0607155} 
\be{ed1}
Q_e^2\equiv Q_e\cdot Q_e = 2 n/N, \qquad Q_m^2 \equiv Q_m\cdot
Q_m = 2 (Q_1- \beta   ), \qquad Q_e\cdot Q_m = J\, ,
\ee
where $\cdot$ denotes T-duality invariant inner product.
The $-\beta$
term in the expression for $Q_m^2$
reflects the fact that a D5-brane
wrapped on $\MM$ carries $-\chi(\MM)/24$ unit of
D1-brane charge.

The S-duality symmetry of the theory
in the asymmetric orbifold description is related to the
global diffeomorphism 
symmetry of the torus $S^1\times \wt S^1$ 
in the original description. More precisely it is
the subgroup of this global diffeomorphism group which 
leaves invariant $2\pi/N$ translation along $S^1$, and is
represented by the $\Gamma_1(N)$ matrices
$\pmatrix{a &b
\cr c & d}$ satisfying
\be{descra}
ad-bc=1, \quad a,d\in 1 + N\ZZZ, \quad c \in N\ZZZ, \quad b\in \ZZZ\, .
\ee
The duality transformation acts on the electric and the magnetic charge
vectors as 
\be{try2}
\pmatrix{Q_e\cr Q_m} \to \pmatrix{a &b
\cr c & d }
\pmatrix{Q_e\cr Q_m}\, .
\ee

Our goal is to find the spectrum  of 1/4 BPS 
states with charge quantum
numbers $(Q_e,Q_m)$. Since these states break 12 of the 16 
supersymmetry generators of the theory, quantization of the fermionic
zero modes associated with the broken supersymmetry generators
gives rise to 
$2^6=64$-fold degeneracy, with equal number of 
bosonic and fermionic states. This 64-fold degeneracy is associated
with the size of the 1/4-BPS supermultiplet, and a generic 1/4 BPS
state is obtained by tensoring the basic supermultiplet containing
64 states with helicity ranging from $-{3\over 2}$ to ${3\over 2}$
with a supersymmetry invariant state
which could be either
bosonic of fermionic. We shall call such supermultiplets bosonic 
and fermionic supermultiplets respectively, and denote by
$d(Q_e,Q_m)$ the number of 1/4 BPS 
bosonic supermultiplets minus the
number of 1/4 BPS 
fermionic supermultiplets for a given set of charges
$(Q_e,Q_m)$.

Another description of $d(Q_e,Q_m)$, equivalent to the one given
above, is as follows\cite{9708062}. 
If $h$ denotes the helicity of a  state, then
\be{ed3}
d(Q_e,Q_m) = {2^6\over 6!}\, Tr \left((-1)^{2h} h^6\right)\, ,
\ee
where the trace is taken over all 1/4 BPS states with charge
quantum numbers $(Q_e,Q_m)$.

In the present example the charges $(Q_e,Q_m)$ are labelled by the
set of integers $Q_1$, $n$ and $J$ together with the 
number of D5-branes along $\MM\times S^1$ and the
number of Kaluza-Klein monopoles associated with the circle
$\wt S^1$ in the original description, both of which have 
been taken to be 1. 
We shall denote by $h(Q_1,n,J)$ the number of bosonic supermultiplets
minus the number of fermionic supermultiplets carrying quantum
numbers $(Q_1,n,J)$. Computation of $h(Q_1,n,J)$ is best done
in the weak coupling limit of the
original description of the system where 
the quantum numbers  $n$ and $J$
arise from three different sources\cite{0605210}: the
excitations of the Kaluza-Klein monopole which can carry
certain amount of momentum $-l_0'$ along $S^1$, the  overall 
motion of the D1-D5 system in the background of the
Kaluza-Klein monopole which can carry certain amount of
momentum $-l_0$ along $S^1$ and $j_0$ along $\wt S^1$ and
the  motion of the D1-branes  in the plane of the
D5-brane carrying  total  momentum $-L$ along $S^1$ and $J'$
along $\wt S^1$. Thus we have
\be{esi1}
  l_0'+l_0+
L = n, \qquad j_0+  J'=J\, .
\ee
Let
\be{esi8}
f(\wt \rho,\wt \sigma,\wt v) = \sum_{Q_1, n, J} h(Q_1,n, J)
e^{2\pi i ( \wt \rho n + \wt \sigma Q_1/N +\wt v J)} \, ,
\ee
denote the partition function of the system. Then in the
weak coupling limit we can ignore the interaction between the
three different sets of degrees of freedom described above, and
$f(\wt \rho,\wt \sigma,\wt v)$ is obtained as a product of
three separate partition functions:
\bea{esi9}
f(\wt \rho,\wt \sigma,\wt v ) &=& {1\over 64} 
\, \sum_{Q_1,L,J'} d_{D1}(Q_1,L,J') 
e^{2\pi i ( 
\wt \sigma Q_1 /N +\wt \rho L + \wt v J')}
\nonumber \\
&& \, \left(\sum_{l_0, j_0} 
d_{CM}(l_0, j_0) e^{2\pi i l_0\wt\rho + 2\pi i
j_0\wt v}\right) \, 
\left(\sum_{l_0'} d_{KK}(l_0') e^{2\pi i l_0' \wt\rho} \right)\, ,
\eea
where $d_{D1}(Q_1,L,J')$ is the degeneracy of $Q_1$ D1-branes
moving in the plane of the D5-brane carrying momenta $(-L,J')$
along $(S^1,\wt S^1)$, $d_{CM}(l_0,j_0)$ is
the  degeneracy associated with the overall
motion of the D1-D5 system
in the background of the Kaluza-Klein monopole carrying
momenta $(-l_0, j_0)$ along $(S^1,\wt S^1)$ and
$d_{KK}(l_0')$ denotes the degeneracy associated with
the excitations of a Kaluza-Klein monopole carrying momentum
$-l_0'$ along $S^1$. The factor of 1/64
in \refb{esi9} accounts for the fact that a single 1/4 BPS
supermultiplet has 64 states. In each of these sectors we count
the degeneracy weighted by $(-1)^F$ with $F$ denoting space-time
fermion number of the state, except for the parts obtained by
quantizing the fermion zero-modes associated with the broken
supersymmetry generators. Since a Kaluza-Klein monopole
in type IIB string theory on $K3\times S^1\times \wt S^1$ breaks 8
of the 16 supersymmetries, quantization of the fermion
zero modes associated with the broken supersymmetry generators give
rise to a 16-fold degeneracy which appears as a factor
in $d_{KK}(l_0')$. Furthermore since a D1-D5 system in the
background of a Kaluza-Klein monopole in type IIB on $K3\times S^1
\times \wt S^1$ breaks 4 of the 8 remaining supersymmetry generators,
we get a 4-fold degeneracy from the associated fermion zero modes
appearing as a factor in $d_{CM}(l_0, j_0)$. This factor of $16\times 4$
cancel the $1/64$ factor in \refb{esi9}. After separating out this
factor, we count the contribution to the degeneracy from the rest
of the degrees of freedom 
weighted by a factor of $(-1)^F$.

We shall now compute each of the three pieces, $d_{KK}(l_0')$,
$d_{CM}(l_0, j_0)$ and $d_{D1}(Q_1,L,J')$ separately.

\subsection{Counting States of the 
Kaluza-Klein Monopole} \label{s1}

We consider type IIB string theory 
in the background 
$\MM\times TN\times S^1$
where $TN$ denotes Taub-NUT space. This describes
type IIB string theory
compactified on 
$\MM\times S^1\times \wt S^1$ in the presence of
a Kaluza-Klein monopole, with $\wt S^1$ identified with the
asymptotic circle of the Taub-NUT space.
We now  
take an
orbifold of the theory by a $\ZZZ_N$ group
generated by the transformation $g$. 
Our goal is to compute the degeneracy
of the half-BPS states of the Kaluza-Klein 
monopole carrying momentum $-l_0'$ along $S^1$.

The world-volume of the Kaluza-Klein monopole is 5+1 
dimensional with the five spatial directions lying along $\MM
\times S^1$. By taking the size of $\MM$ to be much smaller
than that of $S^1$ we shall regard this as a 1+1 dimensional
theory, obtained by dimensional reduction of the original
5+1 dimensional theory on $\MM$. Since the supersymmetry
generators of type IIB string theory
on K3 are chiral, the world-volume supersymmetry on the 
Kaluza-Klein monopole will also be chiral, 
acting on the right-moving
degrees of freedom of the 1+1 dimensional field theory. Thus the
BPS states of the Kaluza-Klein monopole will correspond to
states in this field theory where the right-moving
oscillators are in their ground state. In order to
count these states we first need to determine the low energy limit
of this world-volume theory.
Since a Kaluza-Klein monopole has three transverse
directions,  there are three non-chiral massless
bosonic fields on the world-sheet associated with oscillations
in these transverse directions.
Since
Taub-NUT space has a normalizable self-dual
harmonic 2-form\cite{brill,pope}, we
can get two additional non-chiral scalar 
modes on the world-sheet of the Kaluza-Klein
monopole by reducing the two 2-form fields of type IIB
string theory along
this harmonic
2-form.  Finally,
the self-dual four form field of type IIB theory, 
reduced along the tensor product of the harmonic
2-form on $TN$ and a harmonic 2-form on $\MM$, can give
rise to a chiral scalar field on the world-sheet. The chirality of
the scalar field is correlated with whether the corresponding
harmonic 2-form on $\MM$ is self-dual or anti-self-dual.
This gives 
3 right-moving and $P$ left-moving scalars where $P=3$ for 
$\MM=T^4$ and 19 for $\MM=K3$. Thus we have
altogether 8 right-moving scalar fields 
and $P+5$ left-moving scalar
fields on the world-volume of the Kaluza-Klein monopole.

Next we turn to the spectrum of massless fermions
in this world-volume theory.
These typically arise from broken supersymmetry generators.
Since type IIB string theory on K3 has 16 unbroken 
supersymmetries\footnote{In this section we shall refer
to unbroken supersymmetries in various context. Some time it
may refer to the symmetry of a given compactification, 
and some time it
will refer to  the symmetry of a given brane
configuration.  The reader
must carefully examine the context in which the symmetry is
being discussed, since the number of unbroken generators and
their action on various fields depend crucially on this
information.}
of which 8 are broken in the presence of the Taub-NUT space,
we have 8 fermionic zero modes. Since the supersymmetry generators
in type IIB on K3 are chiral, the fermionic zero modes associated with
broken supersymmetries are also chiral, and are 
right-moving on the 
world-sheet. 
On the other hand if we take type IIB on $T^4$ 
we have altogether 32
unbroken supercharges of which 16 are broken in the presence of the
Taub-NUT space. Since type IIB on $T^4$ is a non-chiral theory,
we have 8 right-moving and 8 left-moving zero modes. 

To summarize, the world-sheet theory describing the dynamics
of the Kaluza-Klein monopole always
contains 8 bosonic and 8  fermionic right-moving modes.
For $\MM=K3$ the world-sheet theory has 24 left-moving
bosonic modes and no left-moving
fermionic modes whereas for $\MM=T^4$ the world-sheet theory
has 8 left-moving bosonic and 8
left-moving fermionic modes. 

We shall now determine the $\wt g$
transformation properties of these modes. 
Since $\wt g$ commutes with the supersymmetries of type IIB on
$K3$, all the right-handed fermions living on the world-sheet
theory, associated with the broken supersymmetry
generators in the presence of Kaluza-Klein
monopole, must be neutral under $\wt g$. 
Since $\wt g$ also commutes with the unbroken
supersymmetry generators which transforms the right-moving
world-sheet fermions into right-moving world-sheet scalars and
vice versa, the 8 right-moving bosons on the world-volume of the
Kaluza-Klein monopole must also be invariant under $\wt g$.
Five of the left-moving
bosons, associated with the  3 transverse  degree of freedom
and the modes of the 2-form fields along the Taub-NUT space are
also invariant under $\wt g$ since
$\wt g$ acts trivially on the Taub-NUT space. 
The action  of $\wt g$ on the other $P$ left-moving
bosons is represented
by its action on the $P$
left-handed 2-forms on $\MM$.  
This completely determines the action of $\wt g$ on all the
$P+5$ left-moving bosons. Since from the analysis of section \ref{s0}
we know that $\wt g$ leaves invariant the harmonic 0-form, 4-form
and all the three right-handed 2-forms on $\MM$, 
we see that  the net action
of $\wt g$ on the $(P+5)$ left-handed bosonic fields on the world-sheet
of the Kaluza-Klein monopole is in one to one
correspondence with the action of $\wt g$
on the $(P+5)$ even degree harmonic forms on $\MM$, consisting of
$P$ left-handed 2-forms, three $\wt g$ invariant
right-handed 2-forms,
a $\wt g$ invariant
0-form and a $\wt g$ invariant
4-form.
 
What remains is to determine the action of $\wt g$ on the left-moving
fermions.
We shall now show that this
can be represented by the action of $\wt g$
on the harmonic 1- and 3-forms of $\MM$. For $\MM=K3$ there
are no 1- or 3-forms and no left-moving fermions on
the world-sheet of the Kaluza-Klein monopole. Hence the result
holds trivially. 
For $\MM=T^4$ there are eight left-moving fermions and eight
right-moving fermions. 
These are associated with the sixteen  supersymmetry
generators which are broken in the presence of
a Kaluza-Klein monopole in type IIB string theory on $T^4
\times S^1\times \wt S^1$, and hence transform in the spinor
representation of the tangent space $SO(4)_\parallel$ 
group associated
with the $T^4$ direction.
Now $\wt g$ is an element of this group describing $2\pi/N$ rotation
in one plane and $-2\pi/N$ rotation in an orthogonal plane.
Translating this into the spinor representation we see that the net
effect is to leave half of the eight fermions invariant, rotate two
pairs of fermions by $2\pi/N$ and rotate the other two pairs of fermions
by $-2\pi/N$. Since we have already seen that the right-moving
fermions are neutral under $\wt g$, the action of $\wt g$ on the
left-moving fermions is to rotate two pairs of fermions by 
$2\pi/N$ and another two pairs of fermions by $-2\pi/N$.
This is identical to the action of
$\wt g$ on the harmonic 1- and 3-forms of $T^4$ given
in \refb{eoneform}
and \refb{ethreeform}.

Thus the problem of studying the $\wt g$ transformation properties
of the left-moving degrees of freedom on the world-sheet 
reduces to the problem of finding
the action of $\wt g$ on the even and odd degree
harmonic forms of $\MM$.
We now map this
problem into an equivalent problem as follows. Let us consider
a (4,4) superconformal $\sigma$-model in (1+1)
dimension with target space $\MM$ as described in section
\ref{s0}
and consider the quantity
\be{e1}
Q_{0,s}
= Tr_{RR}\left ((-1)^{F_L+F_R} \wt g^s e^{2\pi i \tau L_0}
e^{-2\pi i \bar\tau \bar L_0}
\right)\, ,
\ee
with $Q_{r,s}$ defined through \refb{eqrs}. As discussed at the end
of section \ref{s0},
$Q_{0,s}$  counts the difference between the number of even
degree harmonic forms and odd degree harmonic forms, 
weighted
by $\wt g^s$. 
Using the results of our previous analysis this can be rewritten
as
\bea{e3}
Q_{0,s} &=& \hbox{number of left handed 
bosons weighted by $\wt g^s$}
\nonumber \\
&& - \hbox{number of left handed fermions weighted by $\wt g^s$}
\, .
\eea
Let $n_l$ be the number of left-handed bosons minus fermions
carrying $\wt g$ quantum number $e^{2\pi i l/N}$. Then
we have from \refb{e3}
\be{e4}
n_l = {1\over N} \, \sum_{s=0}^{N-1} 
e^{-2\pi i l s/N}
\, Q_{0,s}\, .
\ee
Clearly $n_l$ is invariant under $l\to l+N$.

We now turn to the problem of counting the spectrum of BPS
excitations of the Kaluza-Klein monopole. 
First of all
note that
since there are eight right-moving fermions neutral under $\wt g$,
the zero modes of these fermions are $\ZZZ_N$ invariant.
These eight fermionic zero modes may be regarded as the goldstone
modes associated with broken supersymmetry generators.
Since
type IIB string theory on $\MM\times S^1/\ZZZ_N$ has 
16 supersymmetries, and since a Kaluza-Klein monopole breaks
half of these supersymmetries, we expect precisely eight fermionic
zero modes associated with the broken supersymmetry generators.
Upon quantization this produces a 16-fold degeneracy of states
with equal number of bosonic and fermionic states.
This is the correct degeneracy of a single irreducible
short multiplet representing half BPS states in type IIB string
theory compactified on $\MM\times S^1/\ZZZ_N$, and will eventually
become part of the 64-fold degeneracy of a 1/4 BPS supermultiplet
once we tensor this state with the state of the D1-D5 system.
Since supersymmetry
acts on the right-moving sector of the world-volume theory, 
BPS condition requires that all the non-zero mode
right-moving oscillators are in their ground state.
Thus the 
spectrum of BPS states is obtained by taking the tensor product
of this irreducible 16 dimensional supermultiplet 
with either fermionic or bosonic
excitations involving the left-moving degrees of freedom on the
world-volume of the Kaluza-Klein monopole. We shall denote
by $d_{KK}(l_0')/16$ the degeneracy of states associated with
left-moving oscillator excitations carrying total  
momentum $-l_0'$, weighted by $(-1)^{F_L}$ where $F_L$
denotes the contribution to the space-time fermion number from
the left-moving modes on the world-sheet. Thus $d_{KK}(l_0')$
calculates the total degeneracy of half-BPS states weighted by
$(-1)^{F_L}$. 

In order to calculate $d_{KK}(l_0')$ we need to count the number
of ways the total momentum $-l_0'$ can be distributed among
the different left-moving oscillator excitations, subject to the
requirement of $\ZZZ_N$
invariance. Since a mode carrying momentum $-l$ along
$S^1$ picks up a phase of $e^{-2\pi i l/N}$ under 
$2\pi/N$ translation along $S^1$, it must pick up a phase of 
$e^{2\pi i l/N}$ under $\wt g$. Thus the number of 
left-handed bosonic
minus fermionic modes carrying momentum
$l$ along $S^1$ is given by $n_l$ given in eq.\refb{e4}. The
number $d_{KK}(l_0')/16$ can now be identified as the number
of different ways the total momentum $l_0'$ can be distributed
among different oscillators, there being $n_l$ bosonic minus
fermionic oscillators 
carrying momentum $l$. This gives
\be{e5}
\sum_{l_0'} d_{KK}(l_0') e^{2\pi i \wt\rho l_0'} 
= 16\, e^{2\pi i C\wt\rho}\, \prod_{l=1}^\infty (1 -  
e^{2\pi i l\wt\rho})^{-n_l} \, .
\ee
The constant $C$ represents the $l_0'$ quantum number
of the vacuum of the
Kaluza-Klein monopole when all oscillators are in their ground state.
In order to determine $C$ let us consider the dual asymmetric
orbifold description of the
system where the Kaluza-Klein monopole gets mapped to an elementary
heterotic or type IIA string along $S^1$. If $\wh g$ denotes the
image of $g$ in the asymmetric orbifold description, then since
$\wh g$ involves a translation by $2\pi/N$ along $S^1$, the elementary
string along $S^1$ is in the sector twisted by 
$\wh g$. Since the modes of the
Kaluza-Klein monopole get mapped to the degrees of freedom of the
fundamental heterotic or type IIA string,  there are $n_l$ left moving
bosonic minus fermionic modes which pick up a 
phase of $e^{2\pi i l/N}$ under the action of $\wh g$.
$C$ now represents $N\times$
the contribution
to the ground state $L_0$ eigenvalue 
from all the left-moving oscillators,  -- the multiplicative factor
of $N$ arising due to the fact that in the orbifold theory
the $S^1$ direction has period $2\pi/N$, and hence the world-sheet
$\sigma$ coordinate of the dual fundamental string is to be identified
with $N\times$ the coordinate along $S^1$.
Since
a bosonic and a fermionic mode twisted by a phase of $e^{2\pi i \vp}$
for $0\le \vp\le 1$
gives a contribution of $-{1\over 24}+ {1\over 4} \, \vp\, (1-\vp)$ 
and ${1\over 24}-
{1\over 4} \, \vp\, (1-\vp)$ respectively 
to the ground state $L_0$ eigenvalue,
we
 have\footnote{We are counting the contribution from a mode and
 its complex conjugate separately.}
 \be{ecv1}
 C = -{N\over 24} \sum_{l=0}^{N-1} \, n_l + {N\over 4}\, 
 \sum_{l=0}^{N-1} \, n_l \, {l\over N} \, \left( 1 -{l\over N}\right)\, .
 \ee
 Using the expression for $n_l$ given in \refb{e4} we get
 \be{ecv2}
 C = -{1\over 24} \sum_{s=0}^{N-1}\, Q_{0,s}
\sum_{l=0}^{N-1} \, 
  e^{-2\pi i ls/N}\, +{1\over 4} \, \sum_{s=0}^{N-1}\, Q_{0,s}\, 
  \sum_{l=0}^{N-1} \, {l\over N} \, \left( 1 -{l\over N}\right)
  e^{-2\pi i ls/N}\, .
  \ee
  The sum over $l$ can be performed separately for $s=0$ and $s\ne 0$,
  and yields the answer
\be{ecvalue}
C = -\wt \alpha\, ,
\ee
with $\wt \alpha$ defined as in \refb{enn9d}:
\be{eaivalue}
\wt \alpha=  {1\over 24N} \, Q_{0,0} - {1\over 2N}
\, \sum_{s=1}^{N-1} Q_{0,s}\, {e^{-2\pi i s/N}\over
(1-e^{-2\pi i s/N})^2 }
\, .
\ee
The left-right level matching condition of the dual heterotic string
theory guarantees that $C$ and 
hence $\wt\alpha$ must be an integer.
Using \refb{e4}, \refb{eqrsrev}, \refb{ecvalue} 
we can rewrite \refb{e5} as
\be{e8}
\sum_{l_0'} d_{KK}(l_0') e^{2\pi i \wt\rho l_0'} 
=16\, e^{-2\pi i \wt \alpha\wt\rho}\, \prod_{l=1}^\infty (1 -  
e^{2\pi i l\wt\rho})^{-\sum_{s=0}^{N-1} 
e^{-2\pi i l s/N}
\, \left(c^{(0,s)}_0(0) + 2 c^{(0,s)}_1(-1)\right)}\, .
\ee

\subsection{Counting States Associated with the Overall Motion of the
D1-D5 System} \label{s2}

We shall now turn to the computation of the contribution to the partition
function from the overall motion of the D1-D5 system. This has two
components, -- the center of mass motion of the D1-D5 system along
the Taub-NUT space transverse
to the plane of the D5-brane, and the dynamics of the Wilson lines on
the D5-brane along $\MM$. The first component is present irrespective
of the choice of $\MM$ but the second component exits only if $\MM$
has non-contractible one cycles, \i.e.\ for $\MM=T^4$.

The contribution from the first component is clearly
independent of the choice
of $\MM$ and has been found in \cite{0605210}. 
If $d_{transverse}(l_0, j_0)$ denotes the number of states associated
with the transverse motion of the system, carrying momentum $-l_0$
along $S^1$ and $j_0$ along $\wt S^1$, then we have
\bea{eone1}
&& \sum_{l_0, j_0} d_{transverse}(l_0, j_0) e^{2\pi i l_0\wt\rho + 2\pi i
j_0\wt v} = 4 \, e^{-2\pi i \wt v} \,  (1 - e^{-2\pi i \wt v})^{-2}\,
\nonumber \\
&& \qquad \qquad \prod_{n=1}^\infty \left\{
(1 - e^{2\pi i n N\wt\rho})^4 \, ( 1 - e^{2\pi i n N\wt\rho + 2\pi i
\wt v})^{-2} \,  ( 1 - e^{2\pi i n N\wt\rho - 2\pi i
\wt v})^{-2}\right\}\, .
\eea
The factor of 4 comes from the quantization of the 
right-moving fermionic zero modes\cite{0605210}.
As before, in the counting of states associated with the left-moving
oscillators we include a weight factor of $(-1)^{F_L}$.
In expressing the right hand side of \refb{eone1}
as a series we always expand
the terms inside the product in positive powers of $e^{2\pi i\wrh}$.
However for the $e^{-2\pi i \wt v} (1 - e^{-2\pi i \wt v})^{-2}$
factor we have two choices, -- either expand it in powers
of $e^{-2\pi i \wv}$, or rewrite is as $e^{2\pi i \wt v}
 (1 - e^{2\pi i \wt v})^{-2}$ and expand it in 
 powers of $e^{2\pi i\wv}$. These two different ways of 
 expanding yield different spectrum, and the correct choice
 depends on the angle between the circles $S^1$ 
 and $\wt S^1$\cite{0605210,pope,9912082}. 
 As this angle passes through
 90$^\circ$ the spectrum changes discontinuously.
 
Let us now compute the contribution to the 
partition function from the
dynamics of the Wilson lines for $\MM=T^4$. For this
we can ignore the presence of the Kaluza-Klein monopole and 
the D1-branes, and consider the dynamics of a D5-brane wrapped
on $T^4\times S^1$. Taking the $T^4$ to have small size we
can regard the world-volume theory as (1+1) dimensional.
This has eight bosonic modes associated with four Wilson lines
and four transverse coordinates, but we shall only be interested
in the dynamics of the Wilson lines. Similarly there are eight
non-chiral fermionic modes, but four of these, related to the
transverse bosonic modes by the unbroken 
supersymmetry algebra
that commutes with $\wt g$, have already been accounted for in
the partition function \refb{eone1}. Thus we shall consider only
four of the fermionic modes which are superpartners of the
four Wilson lines under the unbroken supersymmetry algebra.

Now $\wt g$ acts as a rotation by
$2\pi/N$ on one pair of Wilson lines 
and as a rotation by $-2\pi/N$
on the other pair. Since the unbroken supersymmetry algebra
commutes with $\wt g$ and furthermore, its action on the D5-brane
world-volume is non-chiral, 
$\wt g$ must act as rotation by $2\pi/N$ on one
pair of fermions and $-2\pi/N$ on the other pair both in the left
and the right-moving sector.  In order to be $\ZZZ_N$
invariant,  the modes which pick a phase of $e^{2\pi i/N}$
under $\wt g$ must carry momentum along $S^1$ of the form
$Nk-1$ for integer $k$, whereas modes which 
pick a phase of $e^{-2\pi i/N}$
under $\wt g$ must carry momentum along $S^1$ of the form
$Nk+1$ for integer $k$. As a result, both in the left and 
the right-moving sector, we have a pair of bosons and a pair
of fermions carrying $S^1$ momentum of the form $Nk+1$, and
a pair of bosons and a pair of fermions carrying $S^1$
momentum of the form $Nk-1$.

Eventually when we place this in the background of the
Kaluza-Klein monopole, only the supersymmetry associated with
the right-moving modes remain unbroken. 
Thus in order to get a BPS state
of the final supersymmetry algebra we must put all the right-moving
oscillators in their ground state and consider only 
left moving excitations.

In order to calculate the partition
function associated with these modes we also need information about
their $j_0$ quantum numbers. Near the center of Taub-NUT
the $j_0$ quantum number corresponds to the sum of the angular
momenta in the two planes
transverse to the D5-brane. The left- and right-moving bosonic
modes associated with the
Wilson lines are neutral under
rotation in planes transverse to the D5-brane and hence do not
carry any $j_0$ charge.
However the fermions, being in the spinor representation
of the tangent space group of the transverse space, do
carry $j_0$ charge. 
Since the net $j_0$ quantum number is 
given by the sum of
$\pm{1\over 2}$ units of angular 
momentum associated
with the two transverse planes, 
a quarter of the fermions carry $j_0=1$, another quarter of them
carry $j_0=-1$ 
and half of them have $j_0=0$.
In order to determine
which fermions carry $j_0=\pm 1$ 
we note that from the point of view of an asymptotic
observer $j_0$ represents momentum
along $\wt S^1$, and hence must commute with
the final unbroken 
supersymmetry generators. Since
these generators relate the right-moving bosons with $j_0=0$
to the right-moving
fermions, the right-moving fermionic excitations must have $j_0=0$. 
Thus the left-moving fermions must have
$j_0=\pm 1$ and hence can be rotated to each other by an
appropriate element of the tangent space $SO(4)_\perp$ group
transverse to the D5-brane. 
Since rotation along the tangent plane transverse
to the D5-brane commutes with $\wt g$, the two left-moving fermions
carrying $\wt g$ quantum number $e^{2\pi i/N}$ must have $j_0=\pm 1$
and the two left-moving fermions
carrying $\wt g$ quantum number $e^{-2\pi i/N}$ 
must have $j_0=\pm 1$.

To summarize, the left-moving excitations on the D5-brane
world-volume, related by supersymmetry transformation
to the Wilson lines along $T^4$,
consist of four bosonic and four fermionic modes. Two of the four
bosonic modes carry momentum along $S^1$ of the form $Nk+1$
and the other two carry momentum along $S^1$ of the form
$Nk-1$, but neither of them carry any momentum along
$\wt S^1$. On the other hand two of the fermionic modes
carry momentum along $S^1$ of the form $Nk+1$ and $\pm 1$
unit of momentum
along $\wt S^1$, and the other two fermionic modes 
carry momentum along $S^1$ of the form $Nk-1$ and $\pm 1$
unit of momentum
along $\wt S^1$. If $d_{wilson}(l_0, j_0)$ denotes
the number of states associated with these modes carrying  total
momentum
$-l_0$  along $S^1$ and total momentum $j_0$ 
along $\wt S^1$, then
\bea{eone2}
&& \sum_{l_0, j_0} d_{wilson}(l_0, j_0)
e^{2\pi i l_0\wt\rho + 2\pi i
j_0\wt v} \nonumber \\ 
 &=& \prod_{l\in N\zzz+1\atop l>0} 
(1 - e^{2\pi il \wt\rho})^{-2} \, 
\prod_{l\in N\zzz-1\atop l>0} 
(1 - e^{2\pi il \wt\rho})^{-2}
\prod_{l\in N\zzz+1\atop l>0} 
(1 - e^{2\pi il \wt\rho+2\pi i \wv})   \nonumber \\
&&
\prod_{l\in N\zzz+1\atop l>0} 
(1 - e^{2\pi il \wt\rho-2\pi i \wv}) \,
\prod_{l\in N\zzz-1\atop l>0} 
(1 - e^{2\pi il \wt\rho+2\pi i \wv})  \,
\prod_{l\in N\zzz-1\atop l>0} 
(1 - e^{2\pi il \wt\rho-2\pi i \wv})  \, .  
\eea

Using \refb{eck3}, \refb{ect4} 
one can show that the partition function
associated with the overall dynamics of the D1-D5 system, 
given by the product of the contribution \refb{eone1}
from the dynamics of the transverse modes and (in case $\MM=T^4$)
the contribution \refb{eone2} from the dynamics of the Wilson lines
along $T^4$, can be written as
\bea{ecmfin}
&& \sum_{l_0,j_0} d_{CM}(l_0, j_0) e^{2\pi i l_0\wt\rho + 2\pi i
j_0\wt v} =
4\, e^{-2\pi i \wt v}\, \prod_{l=1}^\infty (1 -  
e^{2\pi i l\wt\rho})^{2\sum_{s=0}^{N-1} 
e^{-2\pi i l s/N}
\, c^{(0,s)}_1(-1)} \nonumber \\
&& \quad \prod_{l=1}^\infty (1 -  
e^{2\pi i l\wt\rho+2\pi i \wv})^{-\sum_{s=0}^{N-1} 
e^{-2\pi i l s/N}
\, c^{(0,s)}_1(-1)} \, 
\prod_{l=0}^\infty (1 -
e^{2\pi i l\wt\rho-2\pi i \wv})^{-\sum_{s=0}^{N-1} 
e^{-2\pi i l s/N}
\, c^{(0,s)}_1(-1)} \nonumber \\
\eea
both for $\MM=K3$ and $\MM=T^4$.

\subsection{Counting States Associated with the Relative Motion of
the D1-D5 System} \label{srel}

Finally we turn to the problem of counting states associated with the
motion of the D1-brane in the plane of the D5-brane. This can be
done by following a procedure identical to the one
described in \cite{0605210} (which in turn is a generalization of the
analysis of \cite{9608096}) and
yields the answer:
\be{emulti}
\sum_{Q_1,L,J'} d_{D1}(Q_1,L,J') e^{2\pi i ( 
\wc \sigma Q_1 /N +\wc \rho L + \wc v J')}
= \prod_{w,l,j\in \zzz\atop w>0, l\ge 0}  
\left( 1 - e^{2\pi i (\wc \sigma w / N + \wc \rho l + 
\wc v j)}\right)^{-n(w,l,j)}\, ,
\ee
where
\be{esi7.1}
n(w,l,j)  
= 
\sum_{s=0}^{N-1} e^{-2\pi i sl/N} c_b^{(r,s)}(4lw/N - j^2)\, ,
\qquad \hbox{$r=w$ mod $N$, $b=j$ mod 2}\, .
\ee

\subsection{The Full Partition Function}

Using \refb{esi9}, \refb{e8}, \refb{ecmfin} and \refb{emulti} we now get
\be{esi9ex}
f(\wt \rho,\wt \sigma,\wt v )= e^{-2\pi i (\wt \alpha\wt\rho 
+ \wt v)}
 \prod_{b=0}^1\, \prod_{r=0}^{N-1}
\prod_{k'\in \zzz+{r\over N},l\in \zzz, j\in 2\zzz+b
\atop k',l\ge 0, j<0 \, {\rm for}
\, k'=l=0}
\left( 1 - e^{2\pi i (\wt \sigma k'   + \wt \rho l + \wt v j)}\right)^{
-\sum_{s=0}^{N-1} e^{-2\pi i sl/N } c_b^{(r,s)}(4lk' - j^2)}\, .
 \ee
The $k'=0$ term in the last expression comes from the terms involving
$d_{CM}(l_0, j_0)$ and $d_{KK}(l_0')$.  
 Comparing the right hand side of this equation with the expression
 for $\wt\Phi$ given in \refb{enn9a}
we can rewrite \refb{esi9} as
\be{esi9a}
f(\wt \rho,\wt \sigma,\wt v )
= {e^{2\pi i  \wt \gamma \wt \sigma }
\over \wt\Phi(\wt \rho,\wt \sigma, \wt v)}\, ,
\ee
where, from \refb{enn9d},
\be{ea2new}
\wt \gamma\, N= {1\over 24} \, Q_{0,0}={1\over 24 } \chi(\MM)\, .
\ee
Eq.\refb{esi8} now gives
\be{ehexp}
h(Q_1,n,J) =  {1\over N}\, \int _\CC d\wt\rho \, d\wt\sigma \,
d\wt v \, e^{-2\pi i ( \wt \rho n 
+ \wt \sigma (Q_1-\wt \gamma\, N)/N +\wt v J)}\, {1
\over \wt\Phi(\wt \rho,\wt \sigma, \wt v)}\, ,
\ee
where $\CC$ is a three real dimensional subspace of the
three complex dimensional space labelled by $(\wt\rho,\ws,\wv)$,
given by
\bea{ep2int}
Im\, \wt \rho=M_1, \quad Im \, \wt\sigma = M_2, \quad
Im \, \wt v = M_3, \nonumber \\
 0\le Re\, \wt\rho\le 1, \quad
0\le Re\, \wt\sigma\le N, \quad 0\le Re\,  \wt v\le 1\, .
\een
$M_1$, $M_2$ and $M_3$ are large but fixed 
positive numbers.
Identifying $h(Q_1,n,J)$ with the degeneracy $d(Q_e,Q_m)$,
using \refb{ed1}, and noting that $\beta$ defined
in \refb{ed2} is equal to $\wt \gamma N$ given in \refb{ea2new},
we can rewrite \refb{ehexp} as
\be{egg1}
d(Q_e,Q_m) = {1\over N}\, \int _\CC d\wt\rho \, 
d\wt\sigma \,
d\wt v \, e^{-\pi i ( N\wt \rho Q_e^2
+ \wt \sigma Q_m^2/N +2\wt v Q_e\cdot Q_m)}\, {1
\over \wt\Phi(\wt \rho,\wt \sigma, \wt v)}\, .
\ee

\sectiono{S-Duality Invariance of $d(Q_e,Q_m)$} \label{s3.5}

The proof of S-duality invariance of $d(Q_e,Q_m)$ proceeds as
in \cite{0510147,0607155}. 
As described in \refb{try2}, under
the action of S-duality the electric and magnetic charges transform
to
\bea{e3.1}
&&   Q_e \to    Q_e' = a  \,  Q_e + b \,  Q_m,
   \qquad Q_m\to Q_m' = c \,   Q_e + d \,  Q_m\, , \nonumber \\
&&   \quad ad-bc=1, \, a ,d \in N\ZZZ+1, \, b \in \ZZZ, \, 
   c \in N\ZZZ\, .
\eea
Let us define
\bea{e3.2}
&& \wt\Omega\equiv\pmatrix{\wt \rho & \wt v
    \cr \wt v & \wt\sigma},
    \qquad \wt\Omega'\equiv\pmatrix{\wt \rho' & \wt v'
    \cr \wt v' & \wt\sigma'}
   = (\wt A \wt\Omega + \wt B) (\wt C \wt\Omega +\wt D)^{-1},
  \nonumber \\
  && \qquad \pmatrix{\wt A & \wt B\cr
     \wt C & \wt D}= \pmatrix{\tilde a & -\tilde b & 0 & 0\cr -
     \tilde c &\tilde d & 0 & 0\cr
     0 & 0 &\tilde d &\tilde c\cr 0 & 0 &\tilde b &\tilde a}
     \eea
     where
\be{epm}  
\pmatrix{\tilde a &\tilde b\cr\tilde c &\tilde d} = 
\pmatrix{d  & c /N \cr b  N & a }\, .
\ee
This gives
\bea{e3.3}
   \wt \rho' =\tilde a^2\wt\rho +\tilde b^2\wt\sigma - 2\tilde a
   \tilde b \wt v \, ,
   \nonumber \\
   \wt\sigma' =\tilde c^2\wt\rho +\tilde d^2 \wt\sigma - 2\tilde
   c\tilde d\wt v \, ,
   \nonumber \\
   \wt v' = -\tilde a\tilde c\wt\rho - \tilde b\tilde 
   d \wt\sigma + (\tilde a\tilde d + \tilde b\tilde c)\wt v \, .
\een
Using \refb{e3.1}, \refb{epm},
\refb{e3.3} and the quantization laws of
$Q_e^2$, $Q_m^2$ and $Q_e\cdot Q_m$ one can easily verify that
\be{e3.4}
e^{-\pi i ( N\wt \rho Q_e^2
+ \wt \sigma Q_m^2/N +2\wt v Q_e\cdot Q_m)} = 
e^{-\pi i ( N\wt \rho' Q_e^{\prime 2}
+ \wt \sigma' Q_m^{\prime 2}/N +2\wt v' 
Q_e'\cdot Q_m')}   \, ,
\ee
and
\be{e3.5}
d\wt\rho \, 
d\wt\sigma \,
d\wt v = d\wt\rho' \, 
d\wt\sigma' \,
d\wt v' \, .
\ee
One can also verify that the transformation
described in \refb{e3.2} is an element of the group $\wt G$
under which $\wt\Phi$ transforms as a modular form of weight
$k$.\footnote{This can be seen directly from the fact that under
the transformation \refb{e3.2} the integral $\tI_{r,s,b}$ given
in \refb{defirsl} remains unchanged after a suitable relabelling
of the indices $m_1$, $n_1$, $m_2$, $n_2$. Eq.\refb{enn7} then
tells us that $\wt\Phi(\wt\rho,\wt\sigma,\wt v)$ transforms as a 
modular form of weight $k$ under this transformation.} Since
for the transformation \refb{e3.2}, $\det(\wt C\wt\Omega+\wt D)=1$,
we have
\be{e3.6}
   \wt\Phi(\wt \rho',\wt\sigma',\wt v') = \wt\Phi(
   \wt \rho,\wt\sigma,\wt v)\, .
\ee
Finally we note that under the map \refb{e3.3} the three cycle $\CC$ 
gets mapped to itself up to a
shift that can be removed with the help of the shift symmetries
\be{eshift}
\wt\Phi(\wt\rho,\wt\sigma,\wt v) =
\wt\Phi(\wt\rho+1,\wt\sigma,\wt v) =
\wt\Phi(\wt\rho,\wt\sigma+N,\wt v) =
\wt\Phi(\wt\rho,\wt\sigma,\wt v+1)\, ,
\ee
which are manifest from \refb{enn9a}.
  Thus eqs.\refb{e3.4}-\refb{e3.6} allow us to express
\refb{egg1} as
\be{e3.7}
d(Q_e,Q_m) = {1\over N}\, \int _\CC d\wt\rho' \, 
d\wt\sigma' \,
d\wt v' \, e^{-\pi i ( N\wt \rho' Q_e^{\prime 2}
+ \wt \sigma' Q_m^{\prime 2}/N +2\wt v Q_e'\cdot Q_m')}\, {1
\over \wt\Phi(\wt \rho',\wt \sigma', \wt v')}  = d(Q_e',Q_m') \, .
\ee
This proves invariance of $d(Q_e,Q_m)$ under the S-duality
group $\Gamma_1(N)$.

\sectiono{Statistical Entropy Function} \label{s4}

In this section we shall describe the 
behaviour of $d(Q_e,Q_m)$
for large charges, and also compute
the first order corrections to the
leading asymptotic formula.
Our starting point is the expression \refb{egg1} for
$d(Q_e,Q_m)$:
\be{enn16}
d(Q_e, Q_m) =   {1\over N}
\int_{\CC} d\wc \rho d\wc\sigma d\wc v
{1\over \wt\Phi(\wc \rho, \wc \sigma, \wc v)}
\exp\left[ -i\pi (N\wc \rho Q_e^2 + \wc \sigma Q_m^2/N +
2\wc v Q_e\cdot Q_m)\right]\, .
\ee
Using eqs.\refb{e6na}, \refb{enn11} and the result
\be{ejac}
d\wrh d\ws d\wv = (2v -\rho-\sigma)^{-3} d\rho d\sigma d v\, ,
\ee
we can rewrite \refb{enn16} as
\bea{enn17a}
d(Q_e, Q_m)&=&  {1\over N\, C_1}
\int_{\CC'}  d\rho   d\sigma  
dv \, (2v -\rho-
\sigma)^{-k-3} \, {1\over \wh\Phi(\rho, \sigma,  v)} \nonumber \\
&& \exp\left[ -i\pi \left\{ {v^2 -\rho\sigma\over 2v 
-\rho-\sigma}
Q_m^2 +{1\over 2v -\rho-\sigma} Q_e^2 +{2(v-\rho)\over 2v -\rho-
\sigma} Q_e\cdot Q_m\right\}\right]  \nonumber \\
\eea
where $\CC'$ is the image of $\CC$ under the map \refb{e5n}. 
We can evaluate this integral by first performing the $v$ integral using
Cauchy's formula and then carrying out the $\rho$ and $\sigma$
integrals by saddle point approximation. Following the analysis of
\cite{9607026,0605210} we can show that the 
dominant contribution comes
from the pole at
\be{epole}
\wt\rho\wt\sigma -\wt v^2 +\wt v=0 \quad \i.e.\quad v=0\, .
\ee 
{}From \refb{enn12} we see that contribution from this pole is
given by
\bea{enn17}
d(Q_e, Q_m)&\simeq& 
C_0\, \int_{\CC''}  d\rho   d\sigma  
dv \, v^{-2} \, (2v -\rho-
\sigma)^{-k-3} \, \left(g(\rho) g(\sigma)
\right)^{-1} \nonumber \\
&& \exp\left[ -i\pi \left\{ {v^2 -\rho\sigma\over 2v 
-\rho-\sigma}
Q_m^2 +{1\over 2v -\rho-\sigma} Q_e^2 +{2(v-\rho)\over 2v -\rho-
\sigma} Q_e\cdot Q_m\right\}\right]\, , \nonumber \\
\een
where $\CC''$ is a contour around $v=0$, $C_0$ is a constant and
$g(\rho)$ has been defined in \refb{enn13}.
This integral is exactly of the form given in eq.(4.19) of
\cite{0605210}. Thus
subsequent analysis of this integral can be done
following the procedure of
\cite{0605210}, and we arrive at the result 
that the statistical entropy 
\be{estat}
S_{stat}(Q_e,Q_m)\equiv \ln d(Q_e,Q_m)
\ee
is obtained by extremizing the statistical entropy function
\be{enn18}
-\wt\Gamma_B(\vec\tau) = {\pi\over 2 \tau_{2}} \, |Q_e +\tau Q_m|^2
- \ln g(\tau) -\ln g(-\bar\tau)
- (k+2) \ln (2\tau_{2}) + \hbox{constant} + \OO(Q^{-2})
\ee
with respect to the real and imaginary parts of $\tau$.

\sectiono{Black Hole Entropy Function} \label{sblack}

We now turn to the computation of the entropy of a black hole
carrying  charge quantum numbers $(Q_e,Q_m)$ 
and compare it with the statistical entropy computed in
section \ref{s4}. For this we first need to determine the
effective action governing the dynamics of the theory.
The leading order entropy is determined by
the low energy effective action with two derivative terms.
This is the standard action of  $\NN=4$ supergravity
theories. An important class of four derivative corrections to the
action is the Gauss-Bonnet term.
We shall now turn to the
computation of this term.

The calculation is best carried out in the original
description of the theory
as type IIB string theory
compactified on $(\MM\times S^1\times \wt S^1)/\ZZZ_N$. 
We shall denote by $t=t_1+it_2$ 
and $u=u_1+iu_2$ the
Kahler and complex structure moduli of the torus 
$S^1\times \wt S^1$ with the normalization convention that 
is appropriate for the orbifold theory. Thus for example if
$R_1$ and $R_2$ denote the radii of $\wt S^1$ and $S^1$
measured in the string metric, and if the off-diagonal
components of the metric and the anti-symmetric tensor field
are zero, then we shall take $t_2=R_1 R_2/N$ and 
$u_2=R_2/(R_1 N)$,
taking into account the fact that in the orbifold theory the 
various fields have $\wt g$-twisted boundary condition under
a $2\pi R_2/N$ translation along $S^1$ and
$2\pi R_1$ translation along $\wt S^1$. In the same spirit
we shall choose the
units of momentum along $S^1$ and $\wt S^1$ to be $N/R_2$ and
$1/R_1$ respectively, and unit of winding charge along
$S^1$ and $\wt S^1$ to be $2\pi R_2/N$ and $2\pi R_1$
respectively. Thus for example a one unit of winding charge along
$S^1$ actually represents a twisted sector state, with twist $g$.

It is known that one loop quantum corrections in this theory
give rise to a Gauss-Bonnet contribution to the effective 
Lagrangian density
of the form\cite{9708062}:
\be{es2}
\Delta\LL =  \phi(u,\bar u)\,
\left\{ R_{G\mu\nu\rho\sigma} R_G^{\mu\nu\rho\sigma}
- 4 R_{G\mu\nu} R_G^{\mu\nu}
+ R_G^2
\right\} \, ,
\ee
where $\phi(u,\bar u)$ is a function to be determined. 
Note in particular
that $\phi$ is independent of the Kahler modulus $t$ of
$S^1\times\wt S^1$. The analysis of \cite{9708062} shows that
$\phi(u,\bar u)$ is given by the relation:
\be{eh1}
\p_u \phi(u,\bar u) =  \int_\FF \, {d^2\tau\over \tau_2} \, \p_u B_4\, ,
\ee
where
$B_4$ is defined as follows. Let us consider type IIB string theory
compactified on $(\MM\times S^1\times \wt S^1)/\ZZZ_N$ in the
light-cone gauge Green-Schwarz formulation, denote by 
$Tr^f$ the trace over all states in this theory excluding the 
momentum modes associated with the non-compact directions
(since their effect has been already included in arriving at the
$\tau_2$ factor in \refb{eh1}) and denote by $L_0^f$,
$\bar L_0^f$ the Virasoro generators associated with the left
and the right-moving modes, excluding the contribution from the
momentum modes associated with the non-compact directions.
We also define $F_L^f$, $F_R^f$ to be the contribution to the
space-time fermion numbers from the left and the right-moving
modes on the world-sheet.
In this case
\be{eh2}
B_4 = K \,
Tr^f\left( q^{L_0^f}\bar q^{\bar L_0^f} (-1)^{F_L^f
+F_R^f} h^4\right), \qquad q\equiv e^{2\pi i \tau}\, ,
\ee
where $K$ is a constant to be determined later and $h$ denotes
the total helicity of the state.

The evaluation of the right hand side of \refb{eh2} proceeds as
follows. We first note that 
without the $h^4$ term the answer
will vanish due to the fermion zero mode contribution to the
trace since quantization
of a conjugate pair
of fermion zero modes $(\psi_0, \psi_0^\dagger)$
gives rise to a pair of 
states with opposite 
$(-1)^{F_L^f
+F_R^f}$. This can be avoided if we insert a factor of $h$
in the trace and pick the contribution to $h$ from this
particular conjugate pair of fermions since the two
states have the same $(-1)^{F_L^f
+F_R^f}\, h$ quantum numbers. This can be repeated for every
pair of conjugate fermions. In the present example we 
have
altogether 8 fermion zero modes which are
neutral under the orbifold group $\ZZZ_N$, --
4 from the left-moving sector and 4 from the right-moving
sector.
As a result we need four factors of $h$ to soak up all the fermion
zero modes. Thus in effect we can simplify \refb{eh2} by
expressing it as
\be{eh3}
B_4 = K' Tr^{f\prime}
\left( q^{L_0^f}\bar q^{\bar L_0^f} (-1)^{F_L^f
+F_R^f} \right)
\ee
where $K'$ is a  different normalization constant and the prime in
the trace denotes that we should ignore the effect of fermion zero
modes in evaluating the trace.

Since we are using the Green-Schwarz formulation, the 4 left-moving
and 4 right-moving fermions which are neutral under the orbifold group
$\ZZZ_N$ satisfy periodic boundary condition. Thus the effect of the
non-zero mode oscillators associated with these fermions cancel
against the contribution from the non-zero mode bosonic oscillators
associated with the circles $S^1$ and $\wt S^1$ and the two
non-compact directions. This leads to a further simplification in
which the trace can be taken over only the degrees of freedom
associated with the compact space $\MM$ and the bosonic zero modes
associated with the circles $S^1$ and $\wt S^1$. The latter includes
the quantum numbers $m_1$ and $m_2$ denoting the number of units of
momentum along $\wt S^1$ and $S^1$, and the quantum numbers $n_1$ and
$n_2$ denoting the number of units of winding along $\wt S^1$ and
$S^1$. The units of momentum and winding along the two circles are
chosen according to the convention described earlier. Thus for example
$m_2$ unit of momentum along $S^1$ will correspond to a physical
momentum of $Nm_2/R_2$ in string units. This shows that $m_2$ can be
fractional, being quantized in units of $1/N$.  On the other hand a
sector with $n_2$ unit of winding along $S^1$ describes a fundamental
string of length $2\pi n_2 R_2/N$, and hence this state belongs to a
sector twisted by $g^{n_2}$.\footnote{This picture can be called the
  view from `downstairs'. In contrast if we use parameters and units
  which are natural for the theory before orbifolding, it corresponds
  to the view from `upstairs'.}

In this convention the contributions to $\bar L_0^f$ and $L_0^f$ from
the bosonic zero modes associated with $S^1\times \wt S^1$ are given
by, respectively,
\be{eh4}
{1\over 2} k_R^2 = {1\over 4 t_2 u_2} |-m_1 u + m_2
+ n_1 t + n_2 tu|^2\, ,
\qquad 
{1\over 2} k_L^2 = {1\over 2} k_R^2 + m_1 n_1 + m_2 n_2\, .
\ee
Thus \refb{eh3} may now be rewritten as
\be{eh5}
B_4 = {K' \over N} \, \sum_{r=0}^{N-1} \sum_{s=0}^{N-1}
\sum_{\stackrel{m_1, n_1\in \zzz, m_2\in \zzz/N}{n_2\in N\zzz+r
}}
q^{k_L^2/2} \bar q^{ k_R^2/2} e^{2\pi i m_2 s} 
Tr_{RR, \wt g^r} \left ((-1)^{F_L+F_R} \wt g^s 
q^{L_0} \bar q^{\bar L_0}\right)
\, .
\ee
The sum over $s$ in \refb{eh5} arises from the insertion of the
projection operator ${1\over N}\sum_{s=0}^{N-1}g^s$ in the trace,
while the sum over $r$ represents the sum over various twisted sector
states.  $Tr_{RR;\wt g^r}$ denotes trace over the $\wt g^r$-twisted
sector RR states of the (4,4) superconformal field theory with target
space $\MM$.  As required, the quantum number $n_2$ that determines
the part of $g$-twist along $S^1$ is correlated with the integer $r$
that determines the amount of $g$-twist along $\MM$. The $e^{2\pi i
  m_2 s}$ factor represents part of $g^s$ that acts as translation
along $S^1$ while the action of $g^s$ on $\MM$ is represented by the
operator $\wt g^s$ inserted into the trace.

We now note that the trace part in \refb{eh5} is precisely the
quantity $N\, F^{(r,s)}(\tau, z=0)$ defined in \refb{esi4aint} 
for $q=e^{2\pi i\tau}$.  Thus we can rewrite
\refb{eh5} as
\be{eh6}
B_4 = {K' } \, \sum_{r=0}^{N-1} \sum_{s=0}^{N-1}
\sum_{\stackrel{m_1, n_1\in \zzz, m_2\in \zzz/N}{n_2\in N\zzz+r
}}
q^{k_L^2/2} \bar q^{ k_R^2/2} e^{2\pi i m_2 s} 
F^{(r,s)}(\tau,0)
\, .
\ee

We shall now compare \refb{eh6} with the expression for 
$\hI(\wt\rho,\wt\sigma,
\wt v)$ given in \refb{rwthrint}, \refb{enn6} at $\wt\rho=u$,
$\wt\sigma=t$ and $\wt v=0$. 
In this case $p_R^2$, $p_L^2$ defined in \refb{e7n} reduces
to $k_R^2$ and $k_L^2 + {1\over 2}j^2$ respectively, with
$k_R^2$, $k_L^2$ given in \refb{eh4}. As a result we have
\bea{eh8}
&& \hI(u,t,0) = \int_{\FF} \frac{d^2\tau}{\tau_2}\, 
\sum_ {r, s =0}^{N-1}\sum_{b=0}^1\,
\sum_{\stackrel{m_1, n_1\in \zzz, m_2\in \zzz/N}{n_2\in N\zzz+ r, 
j\in 2\zzz + b}}
q^{k_L^2/2} \bar q^{ k_R^2/2} q^{j^2/4}
e^{2\pi i m_2 s} 
h_b^{(r,s)}(\tau)\nonumber \\
&=& \int_{\FF} \frac{d^2\tau}{\tau_2}\, 
\sum_ {r, s =0}^{N-1}\,
\sum_{\stackrel{m_1, n_1\in \zzz, m_2\in \zzz/N}{n_2\in N\zzz+ r}}
q^{k_L^2/2} \bar q^{ k_R^2/2}  
e^{2\pi i m_2 s} (\vt_3(2\tau,0) h_0(\tau)+\vt_2(2\tau,0) h_1(\tau))
\nonumber \\
&=& \int_{\FF} \frac{d^2\tau}{\tau_2}\, 
\sum_ {r, s =0}^{N-1}\,
\sum_{\stackrel{m_1, n_1\in \zzz, m_2\in \zzz/N}{n_2\in N\zzz+ r}}
q^{k_L^2/2} \bar q^{ k_R^2/2}  
e^{2\pi i m_2 s} F^{(r,s)}(\tau, 0)\, ,
\eea
where in the last step we have used eq.\refb{efhrel}.
Comparing \refb{eh6} with \refb{eh8} we see that
\be{eh7}
\int_\FF {d^2\tau\over \tau_2} B_4
= K'\, \hI(u,t,0)\, .
\ee
Using \refb{enn9} and \refb{enn12} we get
\bea{eh9a}
\int_\FF {d^2\tau\over \tau_2} B_4
&=& -2 \, K' \lim_{v\to 0}
\left(   k \ln t_2 + k \ln u_2 + 2\ln v + 2\ln \bar v
+ \ln g(t) + \ln g(\bar t) \right.\nonumber \\
&&
\left. + \ln g(u) + \ln g(\bar u) \right) + \hbox{constant}\, .
\eea
Naively the right hand side diverges in the $v\to 0$
limit. The origin of this
infinity lies in the fact that {\it a priori} the integral $\hI$ as well
as $\int d^2\tau B_4/\tau_2$ has divergences from integration
over the large $\tau_2$
region which needs to be removed by adding constant terms in the
integrand. The constants which need to be added to the integrand
of $K'\hI$ is different from the one that needs to be added to $B_4$.
Once we take into account this difference the right hand side
of \refb{eh9a} should become finite. In order to achieve this
we shall first
regularize the right hand side of \refb{eh9a}
and then remove the divergent part
by subtraction. Since the original regularization where we add an
additive constant to the integrand is duality invariant, we must regularize
the right hand side of \refb{eh9a} in a duality invariant manner. Now
under
a duality transformation of the form $t\to (at+b)/(ct+d)$, $v$ transforms
to $v/(ct+d)$. Similarly under a duality transformation of the form
$u\to (pu+q)/(ru+s)$, $v$ transforms to $v/(ru+s)$. Thus the
combination $v\bar v / (t_2 u_2)$ is invariant under both types of
duality transformation. This suggests that a natural way to remove the
divergence on the right hand side of \refb{eh9a} is to replace 
$v\bar v / (t_2 u_2)$ by a small constant $\epsilon$ and then remove
the $\ln\epsilon$ pieces by subtraction. This gives
\bea{eh9}
\int_\FF {d^2\tau\over \tau_2} B_4
&=& -2 \, K' \left(   (k+2) \ln t_2 + (k+2) \ln u_2 + \ln g(t) 
+ \ln g(\bar t) \right.\nonumber \\
&& \left. + \ln g(u) + \ln g(\bar u) \right) + \hbox{constant}\, .
\eea
Comparing \refb{eh1} with \refb{eh9} we now get
\be{eh10}
\phi(u,\bar u) = - 2 \, K' \, \left( (k+2) \ln u_2 
+ \ln g(u) + \ln g(\bar u)\right)
+\hbox{constant}\, .
\ee

We now turn to the determination of $K'$. This constant is universal
independent of the specific theory we are analysing. Thus we can find it
by working with the  type IIB string theory compactified on 
$K3\times S^1\times \wt S^1$. In
this case $k=10$ and $g(\tau)=\eta(\tau)^{24}$. 
This matches with the known
answer\cite{9906094,0007195} 
for $\phi(u,\bar u)$ if we choose $K'=1/(128\pi^2)$. 
Thus we have
\be{eh10a}
\phi(u,\bar u) = - {1\over 64\pi^2} \, \left( (k+2) \ln u_2 
+ \ln g(u) + \ln g(\bar u)\right)
+\hbox{constant}\, .
\ee

Under the duality map that relates type IIB string theory on the 
$\ZZZ_N$ orbifold of $\MM\times S^1\times \wt S^1$ to an
asymmetric $\ZZZ_N$
orbifold of heterotic or type IIA string theory on $T^6$, the modulus
$u$ of the original 
type IIB string theory gets related to the axion-dilaton
modulus $\tau=a+iS$ of the final asymmetric orbifold theory. Thus in
this description the Gauss-Bonnet term in the effective 
Lagrangian density
takes the form
\be{eh10-}
\Delta\LL =  \phi(\tau,\bar \tau)\,
\left\{ R_{G\mu\nu\rho\sigma} R_G^{\mu\nu\rho\sigma}
- 4 R_{G\mu\nu} R_G^{\mu\nu}
+ R_G^2
\right\} \, ,
\ee
with
\be{eh10b}
\phi(\tau,\bar \tau) = - {1\over 64\pi^2} \, \left( (k+2) \ln \tau_2 
+ \ln g(\tau) + \ln g(\bar \tau)\right)
+\hbox{constant}\, .
\ee
The effect of this term on the computation of the black hole
entropy was analyzed in \cite{0508042}. The resulting entropy
function, after elimination of all the near horizon parameters
except the axion-dilaton field $\tau$, is 
\be{eh11}
\EE ={\pi\over 2 \tau_{2}} \, |Q_e +\tau Q_m|^2
- \ln g(\tau) -\ln g(-\bar\tau)
- (k+2) \ln (2\tau_{2}) + \hbox{constant} + \OO(Q^{-2})\, .
\ee
The black hole entropy is obtained by extremizing this function
with respect to the real and imaginary parts of $\tau$. Since the
black hole entropy function coincides 
with the statistical entropy function
given in \refb{enn18}, we see that the black hole entropy agrees with
the statistical entropy to this order.

 \end{document}